\documentclass[a4paper,11pt]{article}
\pdfoutput=1 

\usepackage{jheppub} 
\usepackage{multirow}
\usepackage{amsmath}
\usepackage{slashed}
\usepackage{amstext,amssymb}
\usepackage{graphicx}
\usepackage{xspace}
\usepackage{color}
\usepackage{units}

\newcommand{\mathsym}[1]{{}}

\newcommand{\baz}{\begin{array}{cc}}
\newcommand{\bad}{\begin{array}{ccc}}
\newcommand{\ba}{\begin{array}{c}}
\newcommand{\ea}{\end{array}}
\newcommand{\be}{\begin{equation}}
\newcommand{\ee}{\end{equation}}
\newcommand{\bea}{\begin{eqnarray}}
\newcommand{\eea}{\end{eqnarray}}

\newcommand{\bi}{\begin{itemize}}
\newcommand{\ei}{\end{itemize}}
\newcommand{\bmt}{\begin{pmatrix}}
\newcommand{\emt}{\end{pmatrix}}
\newcommand{\bt}{\begin{tabular}}
\newcommand{\et}{\end{tabular}}

\newcommand{\benu}{\begin{enumerate}}
\newcommand{\eenu}{\end{enumerate}}

\newcommand{\mwl}{m_{W_L}}
\newcommand{\mwr}{m_{W_R}}

\newcommand{\muegam}{\mu\to e\gamma}
\newcommand{\mueee}{\mu\to 3e}
\newcommand{\mue}{\mu\to e}

\newcommand{\bav}{\begin{array}{cccc}}

\newcommand{\blue}[1]{{\color{blue} #1 }}
\newcommand{\red}[1]{\textcolor{red}{#1}}

\title{Effect of large light-heavy neutrino mixing and natural type-II seesaw dominance to lepton flavor violation and neutrinoless double beta decay}
\author[a]{Nitali Dash,}
\author[b]{Sudhanwa Patra,}
\author[c]{Prativa Pritimita,}
\author[c]{Urjit A. Yajnik}
\affiliation[a]{DESM (Physics), Regional Institute of Education (NCERT), Bhubaneswar 751022, India}
\affiliation[b]{Department of Physics, Indian Institute of Technology Bhilai, Raipur 492015, India}
\affiliation[c]{Department of Physics, Indian Institute of Technology Bombay, Powai, Mumbai 400076, India}

\emailAdd{nitali.dash@gmail.com}
\emailAdd{sudhanwa@iitbhilai.ac.in}
\emailAdd{prativa@iitb.ac.in}
\emailAdd{yajnik@iitb.ac.in}

\abstract{
We derive the lower bound on absolute scale of lightest neutrino mass for normal 
hierarchy and inverted hierarchy pattern of light neutrinos by studying the new 
physics contributions to charged lepton flavour violations in the framework of 
a TeV scale left-right symmetric model. In the model the fermion sector comprises 
of the usual quarks and leptons plus a fermion singlet per generation and the scalar 
sector consists of isospin doublets, triplets and a bidoublet. The framework allows 
large light-heavy neutrino mixing where the light neutrino mass formula is governed 
by natural type-II seesaw mechanism unlike the generic type-II seesaw dominance which 
assumes suppressed light-heavy neutrino mixing. We demonstrate how sizeable loop-induced 
contribution to light neutrino mass is kept under control such that light neutrino mass 
formula is dominantly explained by type-II seesaw mechanism. We examine the heavy neutrino contributions with large light-heavy neutrino 
mixing to charged lepton flavour violating processes like $\mu \to e \gamma$, $\mu \to 3 e$ 
and $\mu \to e$ conversion inside a nuclei. We present a complementary study between neutrinoless double 
beta decay and charged lepton flavour violation taking into account single beta decay bound, double 
beta decay bound and cosmology bounds on neutrino mass sum. 
}

\begin{document} 
\maketitle
\flushbottom

\section{Introduction}
\label{sec:introduction}
Neutrino oscillation data clearly indicate that neutrinos have small but non-zero masses~\cite{Olive:2016xmw}. A simple theoretical paradigm for the origin of neutrino mass is the {\it seesaw} mechanism, which predicts the Majorana nature of neutrinos; for a review, see e.g. Ref.~\cite{Mohapatra:2005wg}. The existence of right-handed (RH) neutrinos, as required for the type-I seesaw mechanism~\cite{Minkowski:1977sc, Mohapatra:1979ia, Yanagida:1979as, GellMann:1980vs}, or the triplet scalars, as required for the type-II seesaw mechanism~\cite{Magg:1980ut, Schechter:1980gr, Cheng:1980qt, Lazarides:1980nt, Mohapatra:1980yp}, can both be naturally motivated in ultraviolet-complete models of neutrino mass. One such example is the left-right symmetric model (LRSM)~\cite{Pati:1974yy, Mohapatra:1974gc, Senjanovic:1975rk}. In particular, TeV-scale models of left-right symmetry breaking have a number of testable consequences for collider signals in the gauge~\cite{Keung:1983uu,  Ferrari:2000sp,  Schmaltz:2010xr, Nemevsek:2011hz, Chen:2011hc, Chakrabortty:2012pp, Das:2012ii, AguilarSaavedra:2012gf, Han:2012vk, Chen:2013fna, Rizzo:2014xma,  Deppisch:2014zta, Deppisch:2015qwa, Gluza:2015goa, Ng:2015hba, Patra:2015bga,  Dobrescu:2015qna, Brehmer:2015cia, Dev:2015pga,  Coloma:2015una, Deppisch:2015cua,   Dev:2015kca, Mondal:2015zba, Aguilar-Saavedra:2015iew,    Lindner:2016lpp, Lindner:2016lxq, Mitra:2016kov, Anamiati:2016uxp,   Khachatryan:2014dka, Aad:2015xaa, Khachatryan:2016jqo} as well as Higgs sector~\cite{Gunion:1989in, Deshpande:1990ip, Polak:1991vf, Barenboim:2001vu, Azuelos:2004mwa, Jung:2008pz, Bambhaniya:2013wza, Dutta:2014dba, Bambhaniya:2014cia,   Bambhaniya:2015wna, Dev:2016dja, ATLAS:2014kca, CMS:2016cpz, ATLAS:2016pbt}, neutrinoless double beta decay ($0\nu\beta\beta$)~\cite{Mohapatra:1980yp, Mohapatra:1981pm, Picciotto:1982qe, Hirsch:1996qw, Arnold:2010tu, Tello:2010am,  Chakrabortty:2012mh, Nemevsek:2012iq, Patra:2012ur, Awasthi:2013ff, Barry:2013xxa, Dev:2013vxa, Huang:2013kma, Dev:2014xea, Ge:2015yqa, Borah:2015ufa,  Awasthi:2015ota,   Horoi:2015gdv, Bambhaniya:2015ipg, Gu:2015uek, Borah:2016iqd, Awasthi:2016kbk}, low-energy charged lepton flavor violation (cLFV)~\cite{Riazuddin:1981hz, Pal:1983bf, Mohapatra:1992uu, Cirigliano:2004mv, Cirigliano:2004tc, Bajc:2009ft, Tello:2010am, Das:2012ii, Barry:2013xxa, Dev:2013oxa,  Borah:2013lva, Chakrabortty:2015zpm, Awasthi:2015ota, Bambhaniya:2015ipg, Borah:2016iqd, Lindner:2016bgg, Bonilla:2016fqd} and electric dipole moment (EDM)~\cite{Ecker:1983dj, Nieves:1986uk, Frere:1991jt, Nemevsek:2012iq, Dev:2014xea, Maiezza:2014ala}, all of which could together shed  light on some of the unresolved issues in neutrino physics, such as the Dirac vs. Majorana nature, mass hierarchy and absolute mass of the left-handed (LH) neutrinos, and the leptonic $CP$ violation. These results will have far-reaching implications for beyond the Standard Model (SM) physics in general.

In the conventional LRSM, where symmetry breaking is implemented with scalar bidoublets and triplets, the light neutrino mass is governed by {\it both} type-I~\cite{Minkowski:1977sc, Mohapatra:1979ia, Yanagida:1979as, GellMann:1980vs} and type-II~\cite{Magg:1980ut, Schechter:1980gr, Cheng:1980qt, Lazarides:1980nt, Mohapatra:1980yp} seesaw contributions: 
\begin{align}
M_\nu \ = \ -M_DM_R^{-1}M_D^T + M_L \ \equiv \ M_\nu^{\rm I}+M_\nu^{\rm II} \, .
\label{eq:1}
\end{align}
Here $M_D$ is the Dirac neutrino mass induced by the bidoublet vacuum expectation value (VEV), while $M_R$ and $M_L$ are the Majorana masses of the right and left-handed neutrinos respectively, induced by the triplet VEVs. For phenomenological purposes, it is usually assumed that only one of the contributions is dominant for the low-scale LRSM, with observable ramifications for different experiments. For instance, in the type-I seesaw dominance, we assume $M_L\to 0$ and the light neutrino mass crucially depends on the Dirac mass matrix $M_D$, or the light-heavy neutrino mixing. In fact, for exact left-right symmetry, $M_D$ can be expressed in terms of $M_\nu$ and $M_R$~\cite{Nemevsek:2012iq}, and it turns out that all the light-heavy neutrino mixing effects are suppressed for TeV-scale parity restoration. On the other hand, the type-II seesaw dominance can be realized with either $M_D\to 0$ or with very high scale of parity restoration. The advantage here is that the light and heavy neutrino mass matrices are directly proportional to each other, thus leading to a more predictive scenario. In Ref. ~\cite{Goswami:2020loc}, the authors have considered these two types of dominance separately to constrain lightest neutrino mass scale and heavy neutrino masses from neutrinoless double beta decay and LFV in a TeV scale LRSM. But this comes with the cost of losing all the light-heavy neutrino mixing effects on the lepton number and/or flavor violating observables.      

In this paper, we explore an extended scenario for low-scale LRSM~\cite{Pritimita:2016fgr}, whose particle content is such that the light neutrino mass generation is governed by a natural type-II seesaw mechanism, while still allowing for observable light-heavy neutrino mixing effects. This has important and non-trivial phenomenological consequences. The beautiful aspect of this model is that the dominant new physics contributions to cLFV, $0\nu\beta\beta$ and electron EDM processes can be expressed 
in terms of the observed light neutrino oscillation parameters and lightest neutrino mass. As a result of this, we can derive constraints on the lightest neutrino mass 
from the non-observation of these rare processes. We also make a complementary study between the low-energy cLFV and $0\nu\beta\beta$ processes within this scenario. However, in order to highlight the contributions of right-handed heavy neutrino and sterile neutrino to LFV decays and $0\nu\beta\beta$ decay, we have focussed only on the diagrams mediated by them and ignored other possible channels.

The paper is organized as follows;  In Sec-\ref{sec:lrsm}, we recapitulate the basic model framework 
of left-right symmetric theory followed by a discussion on type-II seesaw dominance with large light-heavy neutrino mixing and the condition to achieve it. In Sec-\ref{sec:lfv}, we discuss new physics contributions to relevant LFV processes due to this mixing and in Sec-\ref{sec:numasslfv}, we demonstrate how these contributions to LFV constrain light neutrino masses. In Sec-\ref{sec:0nubb}, we study how the light neutrino masses are constrained by new physics contributions to $0\nu\beta\beta$ decay and in Sec-\ref{sec:lnv-lfv} we do a complementarity study between LNV and LFV decays. We also show the variation of dipole moment of electron with lightest neutrino mass and PMNS phase in that section. In Sec-\ref{sec:muong2} we comment on the recent muon (g-2) anomaly results and summarize our results in Sec-\ref{sec:concl}. 

\section{The model framework of natural type-II seesaw dominance}
\label{sec:lrsm}
\subsection{Natural type-II seesaw dominance and large light-heavy neutrino mixing}
The left-right symmetric theory is based on the gauge group $\mathcal{G}_{LR} \equiv SU(3)_c\times SU(2)_L \times SU(2)_R \times U(1)_{B-L}$~\cite{Mohapatra:1974gc, Pati:1974yy,Senjanovic:1975rk}. The fermion sector comprises of quarks and leptons as follows,
\begin{eqnarray}
& &q_{L}=\begin{pmatrix}u_{L}\\
d_{L}\end{pmatrix}\equiv[3,2,1,{\frac{1}{3}}] \quad , \quad q_{R}=\begin{pmatrix}u_{R}\\
d_{R}\end{pmatrix}\equiv[3,1,2,{\frac{1}{3}}]\,,\nonumber \\
& &\ell_{L}=\begin{pmatrix}\nu_{L}\\
e_{L}\end{pmatrix}\equiv[1,2,1,-1] \quad , \quad \ell_{R}=\begin{pmatrix}\nu_{R}\\
e_{R}\end{pmatrix}\equiv[1,1,2,-1] \,
\end{eqnarray}
The electric charge of individual components are related to the third component of $SU(2)_{L,R}$ gauge groups and the difference between baryon and lepton number.
$$Q=T_{3L}+T_{3R}+\frac{B-L}{2}$$

The fermion mass generation including light neutrino masses crucially depends on how the left-right symmetry breaking happens. The left-right symmetry can be spontaneously broken down to SM gauge group $SU(3)_c\times SU(2)_L\times U(1)_Y$ either by assigning VEV to a scalar doublet $H_R$ or scalar triplet $\Delta_R$, or with the help of both. In case the spontaneous symmetry breaking is done with the help of doublet $H_R$ which is the minimal scenario, Majorana masses can't be generated for neutrinos and thus it becomes less interesting from a phenomenology point of view. However left-right symmetry breaking through $\Delta_R$ generates Majorana masses for both light and heavy neutrinos thereby allowing lepton number violation which can be probed by same-sign dilepton signatures at colliders as well as $0\nu\beta\beta$ decay at low-energy experiments. In this case the neutrino mass generation is governed by type-I plus type-II seesaw mechanism but it gives negligible contribution to left-right mixing. The final step of symmetry breaking occurs with the help of bidoublet $\Phi$ which breaks the electroweak gauge group $SU(2)_L\times U(1)_Y$ to $U(1)_{\rm em}$ theory.

We briefly discuss below how the addition of a sterile neutrino $S_L(1,1,1,0)$ per generation to this type-I plus type-II seesaw scheme results in large left-right mixing and makes the scenario more interesting phenomenologically. The neutral lepton sector of generic left-right symmetric theories contains three active left-handed neutrinos ($\nu_L$) and three right-handed neutrinos ($N_R$) which are their $SU(2)_{R}$ counterparts. We extend the theory only by adding three sterile neutrinos ($S_L$), for the purpose of generating light neutrino mass through natural type-II seesaw term, as the type-I seesaw term gets exactly cancelled out in the presence of $S_L$. More importantly this scenario may lead to new non-standard contributions to neutrinoless double beta decay, lepton flavour violation and the $T$ and $CP$-violating electric dipole moment (EDM) of charged leptons because of the light-heavy neutrino mixing. Even though type-II seesaw dominance is assumed in many left-right models ~\cite{Tello:2010am,Chakrabortty:2012mh,Nemevsek:2012iq,Barry:2013xxa, Dev:2013vxa, Ge:2015yqa} in the context of low energy phenomenology, the light-heavy neutrino mixing is very much suppressed in such cases. Our model differs from these frameworks by naturally getting type-II seesaw term instead of assuming it, along with large light-heavy neutrino mixing. The detailed derivation of neutrino masses and mixing within this natural type-II seesaw and the implications of large light-heavy neutrino mixing to $0\nu\beta\beta$ decay can be found in ref~\cite{Pritimita:2016fgr}.

The structure of mass matrix in the basis $(\nu_L, N^c_R, S_L)$ leading to natural type-II seesaw dominance is given as,
\begin{eqnarray} \label{eq:typeIIb}
&&\mathcal{M} =
\begin{split}
\left[ 
\begin{array}{c | c} 
\begin{array}{c c} 
\blue{M_L} & M_D \\ 
M^T_D & \blue{M_R}
\end{array} & 
\begin{array}{c} 
\red{\bf 0} \\ M
\end{array} \\ 
\hline 
\begin{array}{c c} 
\red{\bf 0}  &\quad  M^T
\end{array} & 
\begin{array}{c} 
\blue{\bf 0}
\end{array} \\
\end{array} 
\right]
\end{split}
\mathop{\xrightarrow{\hspace*{2.5cm}}}^{\blue{M_R \gg M > M_D \gg M_L}}_{} 
\left\{
\begin{array}{c} 
m_\nu = M_L \quad \mbox{(type-II seesaw)} 
\\[0.2cm]
\mbox{light-heavy mixing} \simeq  M_D/M
\\[0.2cm]
m_{S} \simeq M M^{-1}_R M^T\,,\quad m_N = M_R
\end{array} \right.  \nonumber
\end{eqnarray}
where $M_D$ is the Dirac neutrino mass matrix connecting $\nu$ and $N_R$, $M$ is the mixing matrix in the $\nu-S_L$ sector, $M_L$ $(M_R)$ is the Majorana mass matrix for left-handed (right-handed) neutrinos. 


\begin{table}
	\begin{center}
		\begin{tabular} {|c|c|c|c|c|c|}\hline
			& Fields & $SU(3)_c$ & $SU(2)_L$ & $SU(2)_R$ & $B-L$ \\ \hline
			\parbox[t]{2mm}{\multirow{5}{*}{\rotatebox[origin=c]{90}{Fermions}}} & $q_L$ & {\bf 3} & {\bf 2} & {\bf 1} & 1/3\\
			& $q_R$ & {\bf 3} & {\bf 1} & {\bf 2} & 1/3\\
			& $\ell_L$ & {\bf 1} & {\bf 2} & {\bf 1} & $-1$ \\
			& $\ell_R$ & {\bf 1} & {\bf 1} & {\bf 2} & $-1$ \\
			& $S$ & {\bf 1} & {\bf 1} & {\bf 1} & 0\\
			\hline
			\parbox[t]{2mm}{\multirow{5}{*}{\rotatebox[origin=c]{90}{Scalars}}} & $\Phi$ & {\bf 1} & {\bf 2} & {\bf 2} & 0 \\
			& $H_L$ & {\bf 1} & {\bf 2} & {\bf 1} & $-1$ \\
			& $H_R$ & {\bf 1} & {\bf 1} & {\bf 2} &  $-1$ \\
			& $\Delta_L$ & {\bf 1} & {\bf 3} & {\bf 1} & 2 \\
			& $\Delta_R$ & {\bf 1} & {\bf 1} & {\bf 3} &  2 \\
			\hline
		\end{tabular}
	\end{center}
	\caption{Particle content of left-right theories with type-II seesaw dominance.}
	\label{tab:1}
\end{table}

The symmetry breaking steps in our model and the subsequent mass generation for fermions and bosons can be summed up as follows. We use both scalar doublets and scalar triplets for left-right symmetry breaking in order to obtain natural type-II seesaw mechanism even with non-negligible $M_D$. The first step of symmetry breaking happens by assigning VEVs to both Higgs doublet $H_R$ and  Higgs triplet $\Delta_R$. As an immediate result of this symmetry breaking the new gauge bosons $W^\pm_R$, $Z_R$ and right-handed Majorana neutrinos get mass. The next step of symmetry breaking is done with the help of SM Higgs doublet contained in the bidoublet $\Phi$, at the scale $M_Z$. The SM fermions and gauge bosons $W_L$ and $Z$ get their mass at this stage of symmetry breaking. The complete particle spectrum of the model is given in Table~\ref{tab:1}. For fermions we have suppressed the family index, which runs from 1 to 3.  

With these fermions and scalars, the interaction lagrangian for leptons can be written as,
\begin{eqnarray}
-\mathcal{L}_{Yuk} &=& \,\overline{\ell_{L}} \left[Y_3 \Phi + Y_4 \widetilde{\Phi} \right] \ell_R
+ f\, \left[\overline{(\ell_{L})^c} \ell_{L} \Delta_L+\overline{(\ell_{R})^c}\ell_{R}\Delta_R\right] \, \nonumber \\
&&+F\, \overline{(\ell_{R})} H_R S^c_L + F^\prime\, \overline{(\ell_{L})} H_L S_L + \mu_S \overline{S^c_L} S_L\ + \mbox{h.c.}\,.
\label{lepton-interaction}
\end{eqnarray}
The scalars take VEVs as follows, 
\begin{eqnarray}
&&\langle \Phi \rangle = \frac{1}{\sqrt{2}}\begin{pmatrix} v_1  & 0 \\  0  & v_2 \end{pmatrix}\, , \quad 
\langle \Delta_{R} \rangle = \frac{1}{\sqrt{2}} \begin{pmatrix} 0  & 0 \\  v_R  & 0 \end{pmatrix}\, , \quad 
\langle \Delta_{L} \rangle = \frac{1}{\sqrt{2}} \begin{pmatrix} 0  & 0 \\  v_L  & 0 \end{pmatrix}\, , \nonumber \\
&&\langle H_R \rangle = \frac{1}{\sqrt{2}}\begin{pmatrix} 0 \\  u_R \end{pmatrix}\, , \quad 
\langle H_L \rangle = \frac{1}{\sqrt{2}}\begin{pmatrix} 0 \\  0 \end{pmatrix}\, . \quad 
\end{eqnarray}
After spontaneous symmetry breaking with the above assignment of VEVs to Higgs scalars, the resulting mass matrix for neutral leptons in the basis $\left(\nu_L, N^c_R, S_L\right)$ can be written as,
\begin{equation}
\mathcal{M}_\nu= \left( \begin{array}{ccc}
M_L                   & M_D   & 0 \\
M^T_D                 & M_R   & M^T \\
0          & M     & 0
\end{array} \right) \, ,
\label{eqn:numatrix}       
\end{equation}
where $M_D$ is the Dirac neutrino mass matrix connecting $\nu_L$ and $N_R$, $M$ is the mixing matrix in the $N_R-S_L$ sector, 
$M_L$ $(M_R)$ is the Majorana mass matrix for left-handed (right-handed) active neutrinos generated dynamically by non-zero 
VEV of scalar triplet $\Delta_L$ $(\Delta_R)$. 

\subsection{Relation between neutrino masses and mixings}
One of the elegant features of this framework is that it connects heavy neutrinos with light neutrinos by expressing heavy neutrino masses in terms of oscillation parameters. 
The light and heavy neutrino masses,
$M_L= f_L \langle \Delta_L \rangle = f v_L $ $\left( M_R= f_R \langle \Delta_R \rangle = f v_R \right)$ are related as,
\begin{equation}
  m_\nu
=
  M_L
\propto
  M_R \,.
\end{equation}
Also in this set up the right-handed neutrino mixing is fully determined by its left-handed counterpart and thus the right-handed and left-handed PMNS matrices are of the same form,
\begin{equation}
 V^{PMNS}_R=
  V^{PMNS}_L\,.
\label{eq:equality-VLR}
\end{equation}
Our prime goal is to derive a bound on the lightest neutrino mass from new physics contributions to lepton flavour violating prcoesses and the $T$ and $CP$-violating electric dipole moment (EDM) of charged leptons. Before estimating different new physics contributions to lepton flavor violation and lepton number violation like neutrinoless double beta decay, we fix here the involved input model parameters. 
\begin{itemize}
\item \underline{Masses and mixing of light neutrinos:-} We consider absolute value of lightest neutrino mass as a free parameter and express other light neutrino masses in terms of lightest neutrino mass. For normal hierarchy of light neutrinos ($ m_1 \sim m_2 << m_3$), different light neutrino masses are related as,
\begin{eqnarray}
 &&m_1 = \mbox{lightest neutrino mass} \;\qquad  \nonumber \\
 &&m_2 = \sqrt{m_1^2 +\Delta m_{\rm sol}^2}\;\qquad \nonumber \\
 &&m_3 = \sqrt{m_1^2 +\Delta m_{\rm atm}^2 + \Delta m_{\rm sol}^2}\;. \label{eq1:NH_m2_m3}
\end{eqnarray}
Similarly for inverted hierarchy ($m_3 << m_1 \sim m_2$), different light neutrino masses are related as,
\begin{eqnarray}
 &&m_3 = \mbox{lightest neutrino mass} \;\qquad  \nonumber \\
 &&m_1 = \sqrt{m_3^2 +\Delta m_{\rm atm}^2}\;\qquad \nonumber \\
 &&m_2 = \sqrt{m_1^2 +\Delta m_{\rm sol}^2 +\Delta M_{\rm atm}^2 }\;. \label{eq1:IH_m1_m2}
\end{eqnarray}
The leptonic PMNS mixing matrix is parametrized in terms of neutrino mixing angles and phases as,
\bea
U_{\rm {PMNS}}&=& \begin{pmatrix} c_{13}c_{12}&c_{13}s_{12}&s_{13}e^{-i\delta}\\
-c_{23}s_{12}-c_{12}s_{13}s_{23}e^{i\delta}&c_{12}c_{23}-s_{12}s_{13}s_{23}e^{i\delta}&s_{23}c_{13}\\
s_{12}s_{23}-c_{12}c_{23}s_{13}e^{i\delta}&-c_{12}s_{23}-s_{12}s_{13}c_{23}e^{i\delta}&c_{13}c_{23}
\end{pmatrix}\cdot \mbox{P}, \label{PMNS} 
\eea
where the mixing angles are denoted by $s_{ij}=\sin \theta_{ij}$, $c_{ij}=\cos \theta_{ij}$ and the diagonal phase matrix carrying Majorana phases $\alpha$ and $\beta$ is denoted by $\mbox{P}=\mbox{diag}\left(1, e^{i\alpha}, e^{i \beta} \right)$. We vary the Majorana phases from $0\to \pi$. The experimental values of different oscillation parameters for both NH and IH patterns of light neutrinos are presented in Table.\ref{table-osc}. The light neutrino masses are in general diagonalised in terms of unitary mixing matrix $U\equiv U_{\rm PMNS}$ in a basis where charged lepton are already diagonal. 
$$m^{\rm diag}_\nu= U^\dagger_{\rm PMNS} m_\nu U^*_{\rm PMNS} = \mbox{diag}\left(m_1, m_2, m_3 \right)\,,$$
and the physical masses are related to the mass matrix in flavour basis as,
$$m_\nu = U_{\rm PMNS} m^{\rm diag}_\nu U^T_{\rm PMNS}\,.$$

\begin{table}[htb!]
\centering
\begin{tabular}{ccc}
        \hline 
Oscillation Parameters & Within 3$\sigma$ range         & within 3$\sigma$ range  \\
           &   ({\it Schwetz et al.}\cite{GonzalezGarcia:2012sz})    & Gonzalez-Garcia et al 
           (\cite{Gonzalez-Garcia:2014bfa}) \\
        \hline \hline
$\Delta m^2_{\rm {21}} [10^{-5} \mbox{eV}^2]$              & 7.00-8.09     & 7.02 - 8.09   \\
$|\Delta m^2_{\rm {31}}(\mbox{NH})| [10^{-3} \mbox{eV}^2]$ & 2.27-2.69     & 2.317 - 2.607   \\
$|\Delta m^2_{\rm {31}}(\mbox{IH})| [10^{-3} \mbox{eV}^2]$ & 2.24-2.65     & 2.307 - 2.590   \\
\hline
$\sin^2\theta_{s}$                                        & 0.27-0.34    & 0.270 - 0.344  \\
$\sin^2\theta_{a}$                                        & 0.34-0.67    & 0.382 - 0.643  \\
$\sin^2\theta_{r}$                                        & 0.016-0.030  & 0.0186 - 0.0250  \\
        \hline
\end{tabular}
\caption{Neutrino oscillation parameters in 3$\sigma$ range.}
\label{table-osc}
\end{table}
\item \underline{Masses of heavy right-handed neutrinos:-} Under the type-II seesaw dominance scheme the light and heavy neutrino masses can be written as $m_\nu = f \langle \Delta_L\rangle$ and $M_N = f \langle \Delta_R\rangle = (v_R/v_L) m_\nu$ with $f_L =f_R =f$. Since $v_L$ and $v_R$ are constants, the light left-handed and heavy right-handed neutrino masses are diagonalized by the same unitary mixing matrix, i.e $U_{\rm PMNS}$. Thus the physical masses for right-handed neutrinos $M_i$ are related to light neutrino mass eigenvalues $m_i$ as $M_i \propto m_i$. This relation implies that if the light neutrinos are normal hierarchical then the heavy right-handed neutrinos would also be hierarchical in the same manner, i.e. if $m_{1} < m_{2} << m_{3}$ then $M_{N_1} < M_{N_2} << M_{N_3}$.

 Thus, if we fix the largest mass eigenvalue of heavy right-handed neutrino as $M_N=M_{N_3}$ then the other two mass eigenvalues of right-handed neutrinos can be expressed in terms of normal hierarchy (NH) pattern of light neutrino masses as,
\begin{subequations}
\begin{eqnarray}
	M_{N_1} &= \frac{m_1}{m_{3}} M_{N},\text{ NH}, 
\label{eq:N1_massrel-typeII} \\
	M_{N_2} &= \frac{m_2}{m_{3}} M_{N},\text{ NH}.
\label{eq:N1_massrel-typeII} 
\end{eqnarray}
\end{subequations}
and when the largest mass eigenvalue of right-handed neutrino is fixed as $M_N=M_{N_2}$ then the other physical masses for heavy right-handed neutrinos would be related to inverted hierarchy (IH) pattern of light neutrino masses as
\begin{subequations}
\begin{eqnarray}
	M_{N_1} &= \frac{m_1}{m_{2}} M_{N},\text{ IH}, 
\label{eq:N1_massrel-typeII} \\
	M_{N_3} &= \frac{m_3}{m_{2}} M_{N},\text{ IH}.
\label{eq:N3_massrel-typeII} 
\end{eqnarray}
\end{subequations}
\item \underline{Masses of sterile neutrinos:-} The approximate seesaw block diagonalization scheme for type-II seesaw dominance gives mass formulas for sterile neutrinos as $M_{S} = -M M^{-1}_R M^T$. Assuming the matrix $M$ proportional to identity matrix $M=m_S \mbox{diag}\{1,1,1\}$, the physical masses are inversely proportional to heavy right-handed neutrinos and therefore inversely proportional to light neutrino masses, i.e, $M_{S_i} \propto 1/M_{N_i} \propto 1/m_i$. As a result of this relation, when the light neutrinos are normal hierarchical, then the sterile neutrinos would be hierarchical in the inverse way, i.e. when $m_{1} < m_{2} << m_{3}$ then  $M_{S_3} < M_{S_2} << M_{S_1}$. Either we can fix the value of $m_S$ or take the largest sterile neutrino mass as constant value and express other sterile neutrino masses in terms of light neutrino mass eigenvalues.  We fix the largest sterile neutrino mass eigen value as $M_S=M_{S_1}$ and the other sterile neutrino mass eigenvalues can be expressed in terms of normal hierarchy (NH) pattern of light neutrino masses as,
\begin{subequations}
\begin{eqnarray}
	M_{S_2} &= \frac{m_1}{m_{2}} M_{S},\text{ NH}, 
\label{eq:S2_massrel-typeII} \\
	M_{S_3} &= \frac{m_1}{m_{3}} M_{S},\text{ NH},
\label{eq:S3_massrel-typeII} 
\end{eqnarray}
\end{subequations}
Similarly by fixing the largest sterile neutrino mass as $M_S=M_{S_3}$ the physical masses of other sterile neutrinos can be expressed in terms of inverted hierarchy (IH) pattern of light neutrino masses as,
\begin{subequations}
\begin{eqnarray}
	M_{S_1} &= \frac{m_3}{m_{1}} M_{S},\text{ IH}, 
\label{eq:S1_massrel-typeII} \\
	M_{S_2} &= \frac{m_3}{m_{2}} M_{S},\text{ IH}.
\label{eq:S3_massrel-typeII} 
\end{eqnarray}
\end{subequations}
\item \underline{Neutrino Mixings:-} The flavor states $\nu_L$, $N_R$ and $S_L$ are related to their mass eigenstates 
in the following way
\begin{eqnarray}
\begin{pmatrix}
\nu_{L} \\ S_{L} \\ N^c_{R}
\end{pmatrix}_\alpha 
&=&
\begin{pmatrix}
{\mbox V}^{\nu\nu} & {\mbox V}^{\nu{S}} & {\mbox V}^{\nu {N}} \\
{\mbox V}^{S\nu} & {\mbox V}^{SS} & {\mbox V}^{SN} \\
{\mbox V}^{N\nu} & {\mbox V}^{NS} & {\mbox V}^{NN} 
\end{pmatrix}_{\alpha i} 
\begin{pmatrix}
\nu_i \\ S_i \\ N_i
\end{pmatrix}  \nonumber \\
&=& \begin{pmatrix}
      U_{\rm PMNS}           & \frac{1}{m_S} M_D U^*_{\rm PMNS}           & \frac{v_L}{v_R} M_D U^{-1}_{\rm PMNS} {m^{\rm diag.}_{\nu}}^{-1}   \\
      \frac{1}{m_S} M^\dagger_D U_{\rm PMNS} & U^*_{\rm PMNS}             & \frac{v_L}{v_R} m_S U^{-1}_{\rm PMNS} {m^{\rm diag.}_{\nu}}^{-1} \\
      \mathcal{O}             & \frac{v_L}{v_R} m_S U^{-1}_{\rm PMNS} {m^{\rm diag.}_{\nu}}^{-1}  & U_{\rm PMNS} 
      \end{pmatrix}_{\alpha i} 
\begin{pmatrix}
\nu_i \\ S_i \\ N_i
\end{pmatrix}.  \nonumber \\
 \label{eq:numixing}
\end{eqnarray}
We express below only those input model parameters in terms of neutrino oscillation parameters that are required for estimating branching ratios for LFV decays. The individual mixing matrices are expressed in terms of Dirac neutrino mass matrix $M_D$, mixing term $M$ and right-handed Majorana mass term $M_R$ as,
\begin{eqnarray}
&&{\mbox V}^{\nu\nu} = U_{\rm PMNS} \,, \quad  {\mbox V}^{\nu{S}} = \frac{1}{m_S} M_D U^*_{\rm PMNS}\, , \quad 
{\mbox V}^{\nu {N}} = \frac{v_L}{v_R} M_D U^{-1}_{\rm PMNS} {\widehat{m}_{\nu}}^{-1}\,, \nonumber  \\
&&{\mbox V}^{S\nu} = \frac{1}{m_S} M^\dagger_D U_{\rm PMNS}\,, \quad {\mbox V}^{SS} = U^*_{\rm PMNS} \, ,\quad 
{\mbox V}^{SN} = \frac{v_L}{v_R} m_S U^{-1}_{\rm PMNS} {\widehat{m}_{\nu}}^{-1} \,, \nonumber \\
&&{\mbox V}^{N\nu} = \mathcal{O}\,, \quad   {\mbox V}^{NS}=\frac{v_L}{v_R} m_S U^{-1}_{\rm PMNS} {\widehat{m}_{\nu}}^{-1}  \,, \quad  
{\mbox V}^{NN} =U_{\rm PMNS}\,.
\label{eq:mixings}
\end{eqnarray}
For simplification we have considered $M$ to be diagonal and degenerate. In general the Dirac neutrino 
mass matrix $M_D$ is either of up-type quark mass matrix or charged lepton mass matrix. However we have considered an $SO(10)$ GUT motivated structure for $M_D$ including RGE effects as,
\begin{eqnarray}
M_D&=&\left(
\begin{array}{ccc}
0.0111  &  0.0384-0.0103\,i  & 0.038- 0.4433\,i \\
0.0384 +0.0103\,i & 0.29281 & 0.8623+ 0.0002\,i \\
0.038+ 0.4433\,i &0.8623-0.0002\,i & 77.7573
\end{array}
\right)\text{GeV}\,.  \nonumber
\end{eqnarray}
\item \underline{Other model parameters:-} The spontaneous symmetry breaking of left-right symmetric model to SM is done by assigning a non-zero VEV to scalar triplet $\Delta_R$ denoted by $v_R$. The value of $v_R$ decides the masses of right-handed charged gauge bosons $W_R^{\pm}$, doubly charged scalar triplet $\Delta^{++}$, right-handed neutrinos and others. Considering the bound on $v_R > 6$~TeV~\cite{Bertolini:2014sua,Maiezza:2010ic,Zhang:2007da,Beall:1981ze,Bambhaniya:2014cia}, we present below the other input model parameters.
\begin{eqnarray}
&&v_R \geq \mbox{15\,TeV}\,, \quad M_{W_R} \geq \mbox{10\,TeV}\,,
\quad M_{\Delta^{++}} \simeq \mbox{10\,TeV}\,, \quad M_{N} \simeq \mbox{1\,TeV}\,.
\end{eqnarray}
The parameters are chosen in such a way that they not only provide the plot for natural type-II seesaw domiance but also ensure that the contributions from charged scalar triplets and right-handed charged gauge boson $W_R$ are negligible. 
\end{itemize}

\section{Lepton flavour violation in left-right symmetric model}
\label{sec:lfv}
A flavour violating process involving charged leptons has not been observed yet. Many new physics models that discuss lepton flavour violation (LFV) are constrained by muon decay experiments since the current limits on $\tau$ observables are less stringent. For the decay $l_{\alpha}\rightarrow 3l_{\beta}$, the SINDRUM experiment has set a limit of 
$\text{BR}(l_{\alpha}\rightarrow 3l_{\beta})< 10^{-12}$~\cite{Bellgardt:1987du} since a long time, which is expected to improve significantly by the future Mu3e experiment. Similarly for the decay $l_{\alpha}\rightarrow l_{\beta}\gamma$, an impressive bound on its branching ratio 
$\text{BR}(l_{\alpha}\rightarrow l_{\beta}\gamma)< 4.2\times10^{-13}$ is provided by the 
MEG collaboration~\cite{Adam:2013mnn} which will be improved by the upgraded MEG-II. Experiments like Mu2e, COMET, PRIME focus on $\mu\rightarrow e$ conversion that will have a sensitivity ranging from $10^{-14}$ to $10^{-18}$. The present bound and future sensitivity of these lepton flavour violating processes are given in Table.\ref{tab:lfv-expt-bound}.

\begin{table}[h!]
\centering
\begin{tabular}{|c|c|c|c|}
\hline \hline
 LFV Decays              & Present Bound             & Near Future Sensitivity \\ 
 (with Branching Ratios) &                           & at ongoing search experiments \\[2mm]
\hline \hline
$\mbox{Br} \left(\mu \to {e\gamma} \right)$                 &$5.7\times 10^{-13}$         &$6\times 10^{-14}$ \\[2mm]
\hline
$\mbox{Br} \left(\tau \to {e\gamma} \right)$   & $3.3\times 10^{-8}$       &$3\times10^{-9}$\\[2mm]
\hline
$\mbox{Br} \left(\tau \to {\mu\gamma} \right)$   & $4.4   \times 10^{-8}$       & $3 \times 10^{-9}$ \\[2mm]
\hline
$\mbox{Br} \left(\mu \to \mbox{3e} \right)$ & $1.0 \times 10^{-12}$       & $10^{-16}$ \\[2mm]
\hline
$\mbox{Br} \left(\tau \to \mbox{eee} \right)$ & $3.0 \times 10^{-8}$     & $10^{-9}$ \\[2mm]
\hline
$\mbox{Br} \left(\tau \to {\mu\mu\mu} \right)$ & $2.0 \times 10^{-8}$      & $3 \times 10^{-9}$ \\[2mm]
\hline
\end{tabular}
\caption{Branching ratios for different LFV processes and their present experimental bound and future sensitivity values taken from refs~\cite{Adam:2013mnn, Baldini:2013ke,Aubert:2009ag,Aushev:2010bq,Bellgardt:1987du,Blondel:2013ia,Hayasaka:2010np}.}
\label{tab:lfv-expt-bound}
\end{table}

Within canonical seesaw models LFV can be induced via light neutrino mixing. However in such models 
the branching ratios are found to be very much suppressed which are well below the present and planned 
experimental sensitivity.
In left-right symmetric models dominant new contributions to LFV arise when symmetry breaking occurs at few TeV scale. A detailed discussion on LFV within manifest left-right symmetric model can be found in \cite{Barry:2013xxa,Cirigliano:2004mv}. The relevant interaction terms which can mediate the processes in a LRSM are given below.
\begin{itemize}
	\item {\bf Charged-current interactions in the lepton sector:} \\
	The charged-current interactions in the lepton sector within the present model where neutral leptons comprising 
	of $\nu_L, N_R, S_L$ are given by
	\begin{eqnarray}
	\mathcal{L}^{\rm \ell}_{\rm CC} &=& \sum_{\alpha=e, \mu, \tau}
	\bigg[\frac{g_L}{\sqrt{2}}\, \overline{\ell}_{\alpha \,L}\, \gamma_\mu {\nu}_{\alpha \,L}\, W^{\mu}_L 
	+\frac{g_R}{\sqrt{2}}\, \overline{\ell}_{\alpha \,R}\, \gamma_\mu {N}_{\alpha \,R}\, W^{\mu}_R \bigg] + \text{h.c.} 
	\nonumber 
	\label{eqn:ccint-flavor-lepton}
	\end{eqnarray}
	Using masses and mixing relation for neutral leptons,
	\begin{eqnarray}
	&&\nu_{eL}= \mbox{V}^{\nu\nu}_{e\, i}\, \nu_i + \mbox{V}^{\nu\, S}_{e\, i}\, S_i + \mbox{V}^{\nu\, N}_{e\, i}\, N_i, \nonumber \\
	&&N_{eR} = 0\times\,\nu_i +\mbox{V}^{N\, S}_{e\, i}\,S_i +                       \mbox{V}^{NN}_{e\, i}\, N_i
	\end{eqnarray}
	the above CC interaction Lagrangian modifies as
	\begin{eqnarray}
	\mathcal{L}^{\rm \ell,m}_{\rm CC}&=& \frac{g_L}{\sqrt{2}}\,
	\bigg[ \overline{e}_{\,L}\, \gamma_\mu 
	\{\mbox{V}^{\nu\nu}_{e\, i}\, \nu_i + \mbox{V}^{\nu\, S}_{e\, i}\, S_i +
	\mbox{V}^{\nu\, N}_{e\, i}\, N_i \}\, 
	W^{\mu}_L \bigg] +\mbox{h.c.} \nonumber \\
	& &  + \frac{g_R}{\sqrt{2}}\,
	\bigg[ \overline{e}_{\,R}\, \gamma_\mu 
	\{\mbox{V}^{N\, S}_{e\, i}\,S_i +
	\mbox{V}^{NN}_{e\, i}\, N_i \}\, 
	W^{\mu}_R \bigg] + \mbox{h.c.}
	\label{eqn:ccint-mass}
	\end{eqnarray}
	\item {\bf Charged-current interactions in the quark sector:} \\
	The relevant charge-current interaction for quarks are
	\begin{eqnarray}
	\mathcal{L}^{\rm q}_{\rm CC} &=& \bigg[\frac{g_L}{\sqrt{2}}\, \overline{u}_{\,L}\, \gamma_\mu {d}_{\,L}\, W^{\mu}_L 
	+\frac{g_R}{\sqrt{2}}\, \overline{u}_{R}\, \gamma_\mu d_{R}\, W^{\mu}_R \bigg] + \text{h.c.} \nonumber \\
	\label{eqn:ccint-flavor-quark}
	\end{eqnarray}
	\item {\bf Scalar triplet interactions:} \\
	The other relevant terms involving scalar triplets are given by
	\begin{eqnarray}
	{\cal L}_{\Delta^{\pm}_{L}} &=& \frac{\Delta_L^+}{\sqrt{2}}\left[\overline{{\nu_L}^c} f \ell_{L}
	+\overline{{\ell_L}^c} f \nu_L\right] +{\rm h.c.}\,,\\
	{\cal L}_{\Delta^{\pm\pm}_{L,R}} &=& \Delta_{L,R}^{++}\overline{{\ell}^c} f P_{L,R}\ell 
	+ \Delta_{L,R}^{--}\overline{\ell} f^\dagger P_{R,L}{\ell}^c\, . \notag 
	\label{eqn:ccint-flavor-scalartriplet}
	\end{eqnarray}
\end{itemize}

In our model, LFV decays can be mediated by heavy right-handed neutrinos $N_R$, extra sterile neutrinos 
$S_L$, charged scalar triplets $\Delta^{\pm \pm}_{L,R}, \Delta^{\pm}_{L,R}$ and 
gauge bosons $W_{L,R}$ which we classify as follows,
\begin{itemize}
	\item due to purely left-handed currents (LL) arising from i) exchange of light neutrinos, 
	ii) exchange of heavy right-handed neutrinos and extra sterile neutrinos with the involvement 
	of light-heavy neutrino mixing, iii) exchange of scalar triplets; 
	\item due to purely right-handed currents (RR) via exchange of right-handed charged gauge boson, right-handed 
	scalar triplet and heavy neutrinos; 
	\item due to involvement of left-handed as well as right-handed currents (LR).
\end{itemize}

 However we focus here only on those contributions which involve large active-sterile neutrino mixing, 
 i.e. due to the neutrinos $N_R$ and $S_L$ in order to constrain light neutrino masses from LFV decays. We ignore the other possible contributions by imposing the following limiting conditions;
\begin{itemize}
	\item The mass of right-handed gauge boson $W_R$ is assumed to be much heavier than the SM gauge boson $W_L$, i.e, $M_{W_R} \gg M_{W_L}$. 
	The mass eigenvalues for charged gauge bosons are given by
	\begin{eqnarray}
	&&M_{W_1}^2 \approx \frac{1}{4} g_L^2 v^2\,, \quad 
	M_{W_2}^2 \approx \frac{1}{2} g_R^2  v^2_R\,.
	\end{eqnarray}
	With $v_R\simeq 15$~TeV and $g_L=g_R\simeq 0.623$, the $W_L-W_R$ mixing is found to be
	\begin{equation}
	|\tan\, 2\xi | \simeq \frac{v_1 v_2}{v^2_R} \propto \frac{M^2_{W_L}}{M^2_{W_R}} \leq 10^{-4}\,.
	\end{equation}
	The mass of $W_R$ is found to be around $7~$TeV leading to negligible contributions to 
	LFV processes via right-handed currents and mixed currents. 
	\item The masses of scalar triplets and other scalars contained in the bidoublet are assumed 
	to be larger than heavy neutrinos, i.e, $M_{\Delta_{L,R}}, M_{H_1} \gg M_{N,S}$. 
\end{itemize}  

\begin{figure}[h!]
	\centering
	\includegraphics[width=0.9\textwidth]{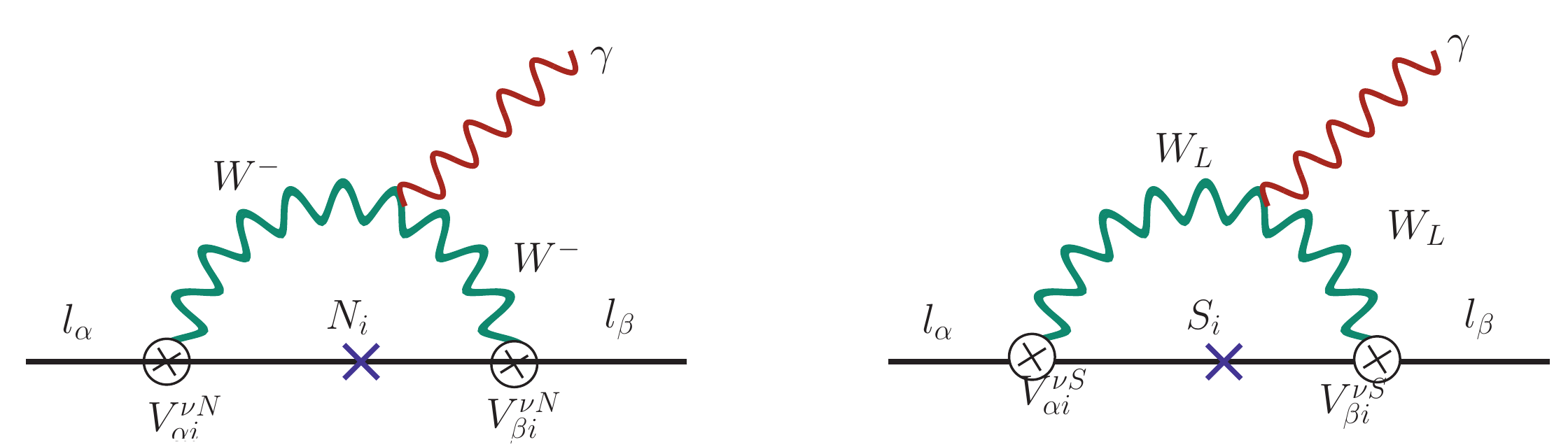}
	\caption{Feynman diagram for lepton flavor violating process $\muegam$ due to the exchange of mass eigenstates of heavy neutrinos $N_i$ and $S_i$.}
	\label{fig:lfv-NS}
\end{figure}
\section{Constraints on light neutrino masses from LFV}
\label{sec:numasslfv}
In this section we derive constraints on absolute scale of lightest neutrino mass inluding both normal hierarchy (NH) and inverted hierarchy (IH) patterns from charged lepton flavour violating processes like $\mu \to e \gamma$, $\mu \to 3e$ and $\mu \to e$ conversion inside a nucleus. We do so by considering the new contributions to these decays due to large light-heavy neutrino mixing in our model. 

The general expression for such a decay can be written as,
\begin{eqnarray*}
&&\Gamma^{(0)}_{\mu}\equiv \Gamma_\nu (\mu^+ \to e^+ \nu_e \overline{\nu_\mu}) \\ 
&&\Gamma^{\rm Z}_{\rm capt.}\equiv \Gamma\left(\mu^{-} + A(Z,N) \rightarrow \nu_\mu + A(Z-1,N+1)\right)
\end{eqnarray*}
and the expressions for branching ratios as,
\begin{eqnarray}
&&\text{Br}_{\mu\rightarrow e\gamma} \equiv \frac{\Gamma(\mu\rightarrow e \gamma)}{\Gamma^{(0)}_{\mu}} \\
&&\text{R}_{\mu \rightarrow e}^A \equiv \frac{\Gamma \left(\mu + A(N,Z)\rightarrow e + A(N,Z)\right)}{\Gamma^{\rm Z}_{\rm capt.}} \\ 
&&\text{Br}_{\mu \rightarrow 3e}\equiv \frac{\Gamma(\mu \rightarrow 3e)}{\Gamma^{(0)}_{\mu}} \, .
\end{eqnarray}

\subsection{Bound on light neutrino mass from $\muegam$}
The effective Lagrangian relevant for the lepton flavor violating process $\mu \to e \gamma$ in our present work can be expressed as~\cite{Barry:2013xxa,Cirigliano:2004mv},
\begin{equation}
\begin{split}
{\cal L}_{\mue} = &-
\frac{e g^2}{4(4\pi)^2\mwl^2}m_\mu\overline{e}\sigma_{\mu\nu}(\mathcal{G}^\gamma_LP_L+\mathcal{G}^\gamma_RP_R)\mu F^{\mu\nu} \\ 
&-\frac{\alpha_W^2}{2\mwl^2}\sum_q \left\{\overline{e}\gamma_\mu\left[\mathcal{W}^q_LP_L
+\mathcal{W}^q_RP_R\right]\mu\; \overline{q}\gamma^\mu q\right\} + {\rm h.c.},
\end{split}
\end{equation}
where $\sigma_{\mu\nu} \equiv \frac{i}{2}[\gamma_\mu,\gamma_\nu]$ and the form factors 
are $\mathcal{G}^\gamma_{L,R}$ and $\mathcal{W}^{u,d}_{L,R}$. The relevant contributions 
to $\muegam$ is given by
\begin{equation}
\begin{split}
i{\cal M}(\muegam) &= \frac{e\alpha_W}{8\pi\mwl^2}\epsilon_\gamma^\mu \overline{e} 
\left[\left(q^2\gamma_\mu - q_\mu\slashed{q}\right)\left(\mathcal{F}^\gamma_L P_L+\mathcal{F}^\gamma_R P_R\right) \right. \\ 
& \left. -im_\mu\sigma_{\mu \nu}q^\nu\left(\mathcal{G}^\gamma_L P_L+\mathcal{G}^\gamma_R P_R\right)\right]\mu,
\end{split} 
\end{equation}
where the anapole and dipole form factors $\mathcal{F}^{\gamma}_{L,R}$ and 
$\mathcal{G}^\gamma_{L,R}$ will be defined in subsequent paragraph. 
\begin{figure}[h!]
	\centering
	\includegraphics[width=0.49\textwidth]{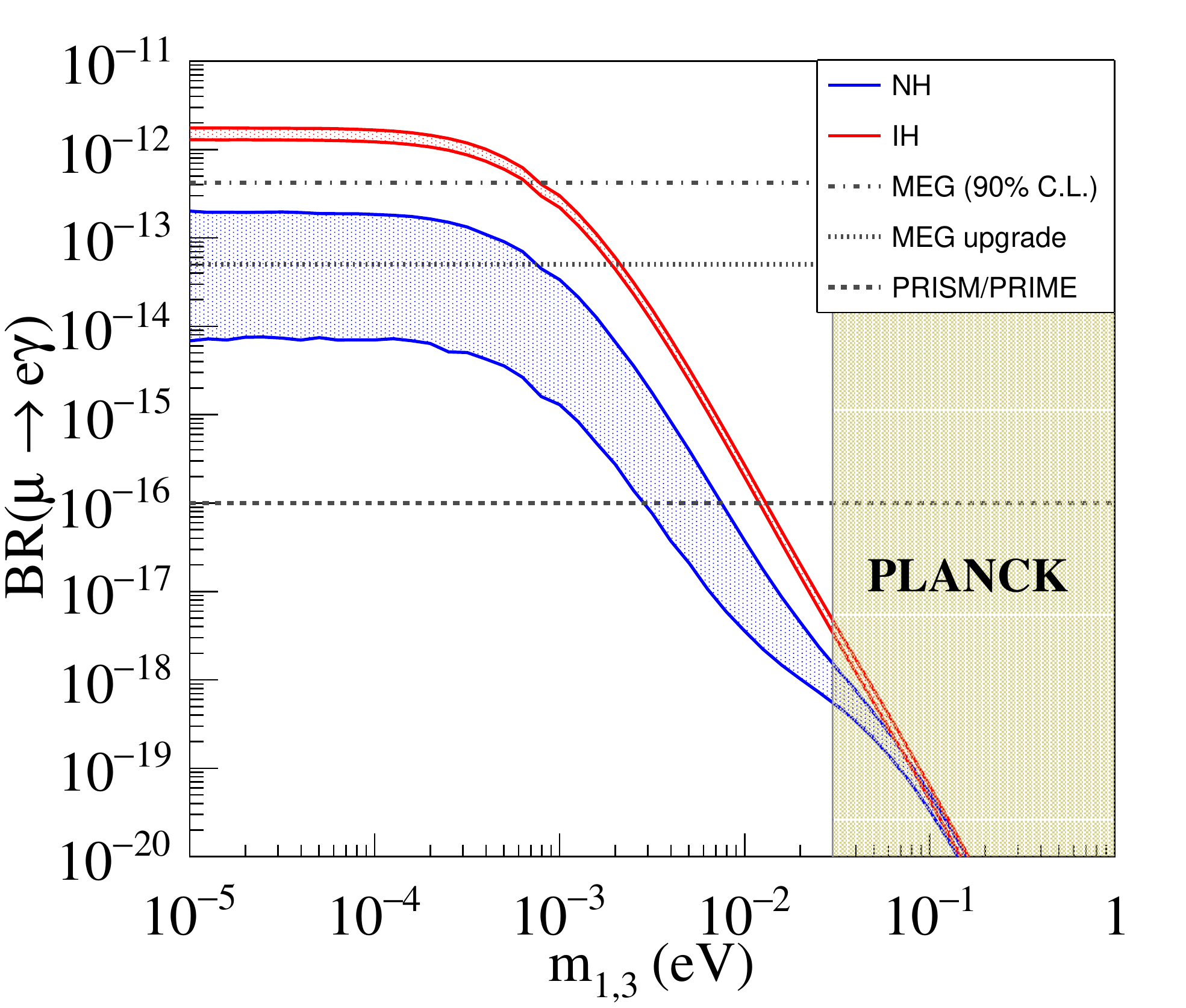}
	\includegraphics[width=0.49\textwidth]{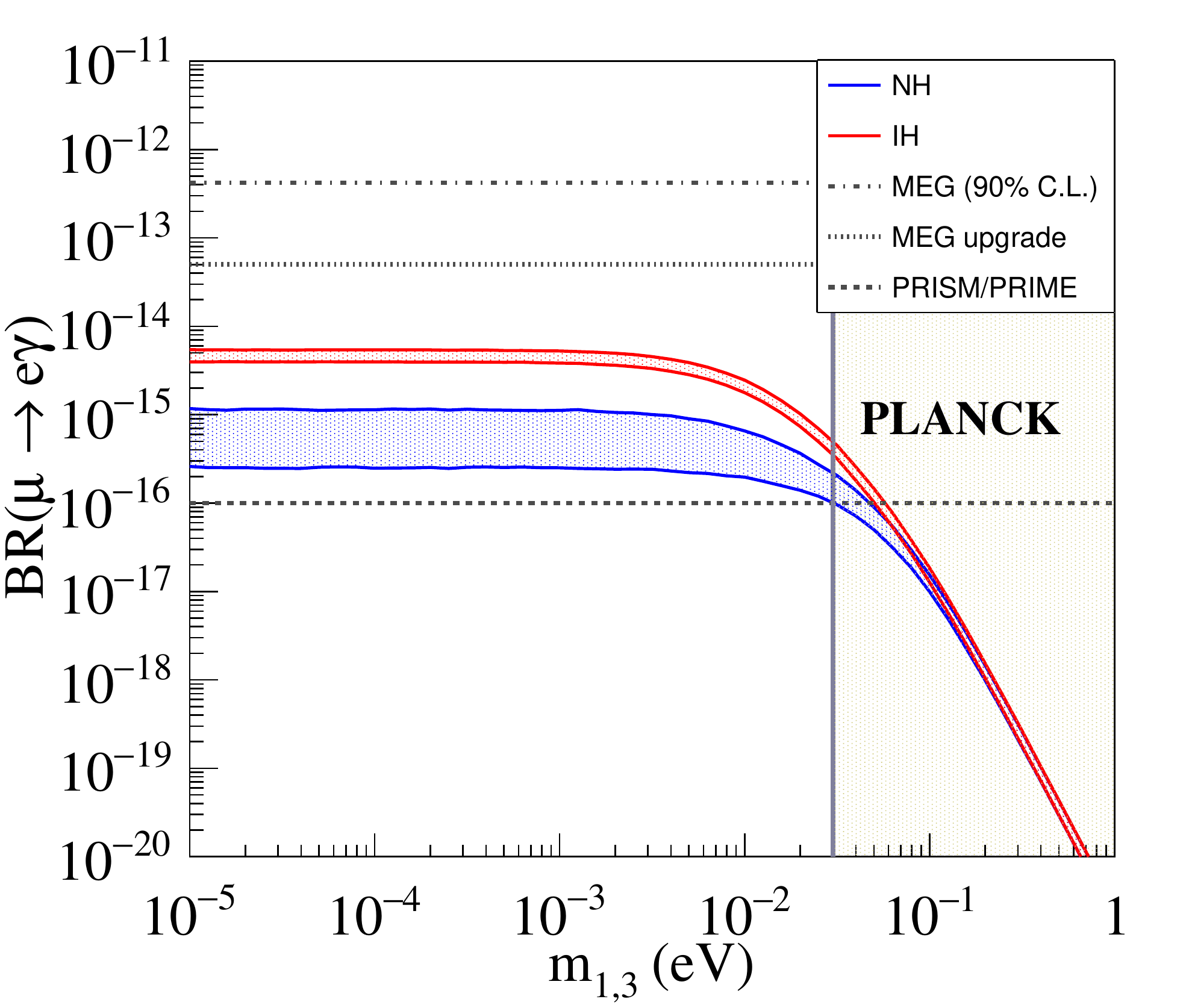}
	\caption{Branching ratio of the lepton flavor violating process, $\mu \to e \gamma$ as a function of lightest neutrino 
		mass $m_1$ for NH and $m_3$ for IH. The blue (NH) and red (IH) coloured regions display the allow due to the exchange of heavy right handed neutrino ($N_R$) (left-panel) and heavy sterile neutrino ($S_L$) (right-panel).}
	\label{fig:mutoegamma}
\end{figure}

The analytic expression for the branching ratio for the lepton flavor violating process $\muegam$ shown in Feynman diagram in Fig.\ref{fig:lfv-NS} due to mediation of heavy neutrinos
\footnote{We have neglected 
	the other contributions to LFV processes due to right-handed gauge boson $W_R$ and $W_L-W_R$ mixing so that we 
	can quantify the effect of light-heavy neutrino mixing on $\text{Br}_{\muegam}$. Even though there are sizable 
	contributions from scalar triplets \cite{Cirigliano:2004mv,Leontaris:1985qc,Swartz:1989qz,Cirigliano:2004tc} we 
	have not included them in our analysis by keeping very large mass scalar triplets. }($N$ and $S$) is given by
\begin{equation}
\text{Br}_{\muegam} = \frac{\alpha_W^3s_W^2}{256\pi^2}\frac{m_\mu^4}{M_{W_L}^4}\frac{m_\mu}{\Gamma_\mu}|\mathcal{G}_\gamma^{\mu e}|^2\,, \label{mueg}
\end{equation}
where $\alpha_W=1/29.0$ as weak fine struture constant, $m_\mu=105$~MeV being the muon mass, $M_{W_L}$ is the SM $W$-boson mass, $s_W\equiv \sin\theta_W$ in which $\theta_W$ is the weak mixing angle and $\Gamma_\mu=2.996\times 10^{-19}$ GeV~\cite{Olive:2016xmw} is the total decay width of the muon. The important parameter,$G_\gamma^{\mu e}$, for deriving constraints on light neutrino masses, is of the following form, 
\begin{eqnarray}
G_\gamma^{\mu e} &=& \bigg| \sum_{i=1}^3 \bigg\{{\mbox{V}^{\nu N}_{\mu i}}^* {\mbox{V}^{\nu N}_{e i}}  \mathcal{G}_{\gamma} \left(x_{N_i}\right) 
+ {\mbox{V}^{\nu S}_{\mu i}}^* {\mbox{V}^{\nu S}_{e i}} \mathcal{G}_{\gamma} \left(x_{S_i}\right)  \bigg\}  \bigg|^2, 
\label{eq:brmuegam_RR_full}
\label{gmue}
\end{eqnarray}
where $x_{N_i}=m_{N_i}^2/M_{W_L}^2$, $x_{S_i}=m_{S_i}^2/M_{W_L}^2$ and the form of loop function is given by
\begin{eqnarray}
G_\gamma(x) &=& -\frac{x(2x^2+5x-1)}{4(1-x)^3}-\frac{3x^3}{2(1-x)^4}\ln x \, .
\end{eqnarray}
The form of loop function $\mathcal{G}_\gamma (x)$ and its dependance with change of lightest neutrino mass $m_1$ (for NH) and $m_3$ (for IH) is presented in appendix-\,\ref{sec:form_loop}. 
The other parameters $\mbox{V}^{\nu N}$ and $\mbox{V}^{\nu S}$ are mixing matrices representing mixing between light-active neutrinos with $N_R$ and $S_L$, respectively.

\par
\begin{figure}[htb!]
	\centering
	\includegraphics[width=0.50\textwidth]{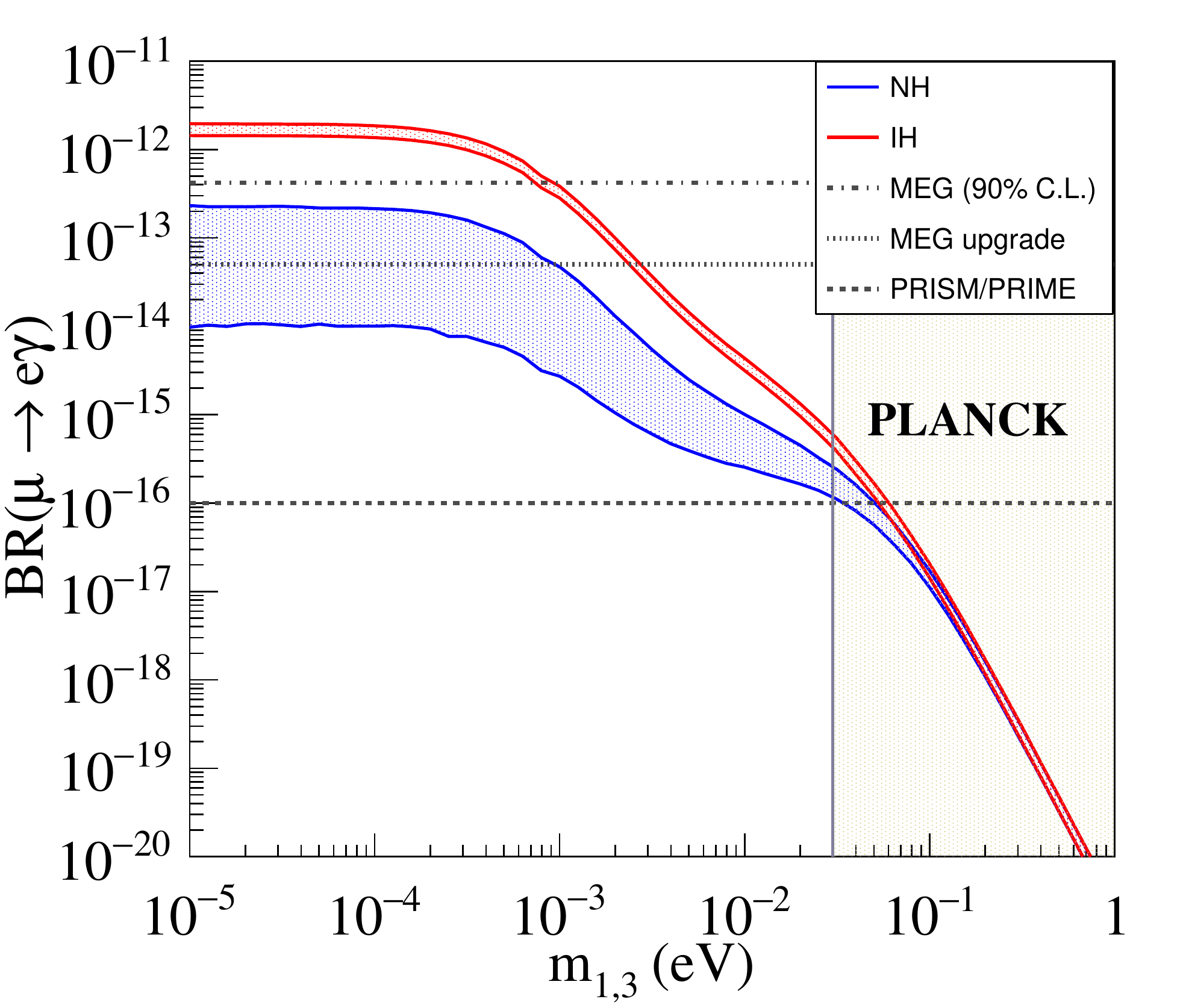}
	\caption{Branching ratio of the process $\mu \to e \gamma$ vs lightest neutrino mass, $m_1$ ($m_3$) for NH (IH) due to 
		the combined effect of heavy right handed  and sterile neutrino.}
	\label{fig:mutoegammaNS}
\end{figure}

 The variation of $\text{Br}_{\muegam}$ as a function of lightest neutrino mass is displayed in Fig.\ref{fig:mutoegamma} with contributions coming from purely $N_R$ is presented in left-panel and for $S_L$ contributions is shown right-panel while the combine contributing is presented in Fig.\ref{fig:mutoegammaNS}. The blue color (NH) and red color (IH) regions are model prediction on $\mu \to e \gamma$ within $3-\sigma$ allowed range of neutrino mixing angles as well as for mass squared differences. In x-axis, $m_1$ for NH ($m_3$ for IH) represents absolute value of light neutrino mass. As given in Table.\ref{tab:lfv-expt-bound}, the current experimental limit and future sensitivity by MEG ($\text{Br}_{\muegam} \leq 4.2\times 10^{-13}$), MEG upgrade ($\text{Br}_{\muegam} \leq 5.0\times 10^{-14}$) and PRISM/PRIME ($\text{Br}_{\muegam} \leq 1.0\times 10^{-16}$) on branching ratio for $\mu \to e \gamma$ are presented in dashed horizontal lines. The vertical shaded regions are PLANCK bound on lightest neutrino mass with 95$\%$ C.L..

The estimated values of $\text{Br}_{\muegam}$ is mostly depend on the sum of the heavy-light neutrino mixing parameters like ${\mbox{V}^{\nu S}_{\mu i}}^* {\mbox{V}^{\nu S}_{e i}}$ for $S_L$ mediated contribution and  ${\mbox{V}^{\nu N}_{\mu i}}^* {\mbox{V}^{\nu N}_{e i}}$ for $N_R$ mediated contribution. The band strucure of NH and IH scenarios are coming since we have taken $3\sigma$ range of oscillation parameters presented in Table.\ref{table-osc} as well as vary the phases; $\delta$ between [$0,2\pi$] and $\alpha, \beta$ between [$0,\pi$]. As presented in Table.\ref{tab:lfv-expt-bound}, the current experimental limit on 
$\text{Br}_{\muegam}$ from MEG~\cite{Adam:2013mnn} ($<4.2 \times 10^{-13}$) and the future experimental sensitivity from MEG Upgrade~\cite{Baldini:2013ke} ($< 5.0 \times 10^{-14}\,$) $\&$ PRISM/PRIME~\cite{Kuno:2005mm} ($<1.0 \times 10^{-16}\,$) are satisfied by the predicted 
branching ratio and one can derive bound on lightest neutrino masses for NH as well as IH case.
\begin{itemize}
	\item Saturating the current experimental MEG bound ($\text{Br}_{\muegam} \leq 4.2\times 10^{-13}$), one can derive the limit on lightest neutrino mass in the range of meV scale in IH case while most of the parameters are ruled out for NH case. 
	\item The projected experimental sensitivity from MEG Upgrade~\cite{Baldini:2013ke} ($< 5.0 \times 10^{-14}\,$) limits lightest neutrino mass less than few meV for NH while $4$~meV for IH case. However, most of the paremeter spaces satisfying PRISM/PRIME~\cite{Kuno:2005mm} ($<1.0 \times 10^{-16}\,$) bound are lying in the quasi-degenerate (QD) pattern of light neutrino masses which already ruled out by PLANCK data.
	
	The predicted branching ratio is allowed by the future sensitivity for lightest neutrino mass $m_1 \leq 10^{-3}$~eV in case of $N_R$ exchange (left-panel Fig.\ref{fig:mutoegamma}) and for interference term (Fig.\ref{fig:mutoegammaNS}). But in case of $S_L$ exchange (right-panel Fig.\ref{fig:mutoegamma}) small range of lightest neutrino mass ($m_1,m_3$) is allowed by future limit.
\end{itemize}


\subsection{Bound on light neutrino mass from $\mu \to 3e$ and $\mu \to e$ conversion}
Under the assumptions that $M_{W_R} \gg M_{W_L}$, $\tan \xi \to 0$ and 
for heavy scalar masses, the only relevant terms contributing to $\mueee$ as a result of 
light-heavy neutrino mixing is given by
\begin{eqnarray}
\text{Br}^{}_{\mueee}&=&
\frac{\alpha_W^4 m_\mu^5}{24576\pi^3\mwl^4\Gamma_\mu} 
\bigg\{ 2\bigg[\bigg| \frac{1}{2} \mathcal{B}^{\mu 3e}_{LL} + 
\mathcal{F}^{Z_1}_L - 2s_W^2(\mathcal{F}^{Z_1}_L-\mathcal{F}^\gamma_L) \bigg|^2 +\bigg|\frac{1}{2} \mathcal{B}^{\mu 3e}_{RR} \bigg|^2 \bigg]
\nonumber \\
&&\hspace*{2cm}+ \bigg|2s_W^2(\mathcal{F}^{Z_1}_L - \mathcal{F}^\gamma_L) \bigg|^2
+8 s_W^2 \mbox{Re}\left(2 \mathcal{F}^{Z_1}_L+\mathcal{B}^{\mu 3e}_{LL} \right) {\mathcal{G}^\gamma_R}^* \nonumber \\
&&\hspace*{2cm} -48 s_W^2 \mbox{Re} \left(\mathcal{F}^{Z_1}_L-\mathcal{F}^\gamma_L \right) {\mathcal{G}^\gamma_R}^*
+32 s_W^4 |\mathcal{G}^\gamma_R|^2 \left[\ln\frac{m^2_\mu}{m^2_e} -\frac{11}{4} \right] \bigg\}\,
\end{eqnarray}
where $m_e (m_\mu)$ is mass of electron (muon), $s_W^2=\sin^2\theta_W$ and other loop factors are 
presented in appendix-\ref{sec:form_loop}. The estimated branching ratio for the process $\mueee$ is satisfied by the experimental limit SINDRUM~\cite{Bellgardt:1987du} ($<10^{-12}$) for a broad range of lightest neutrino mass $m_1,m_3 \leq 0.03$~eV. 

\begin{figure}[h]
	\centering
	\includegraphics[width=0.49\textwidth]{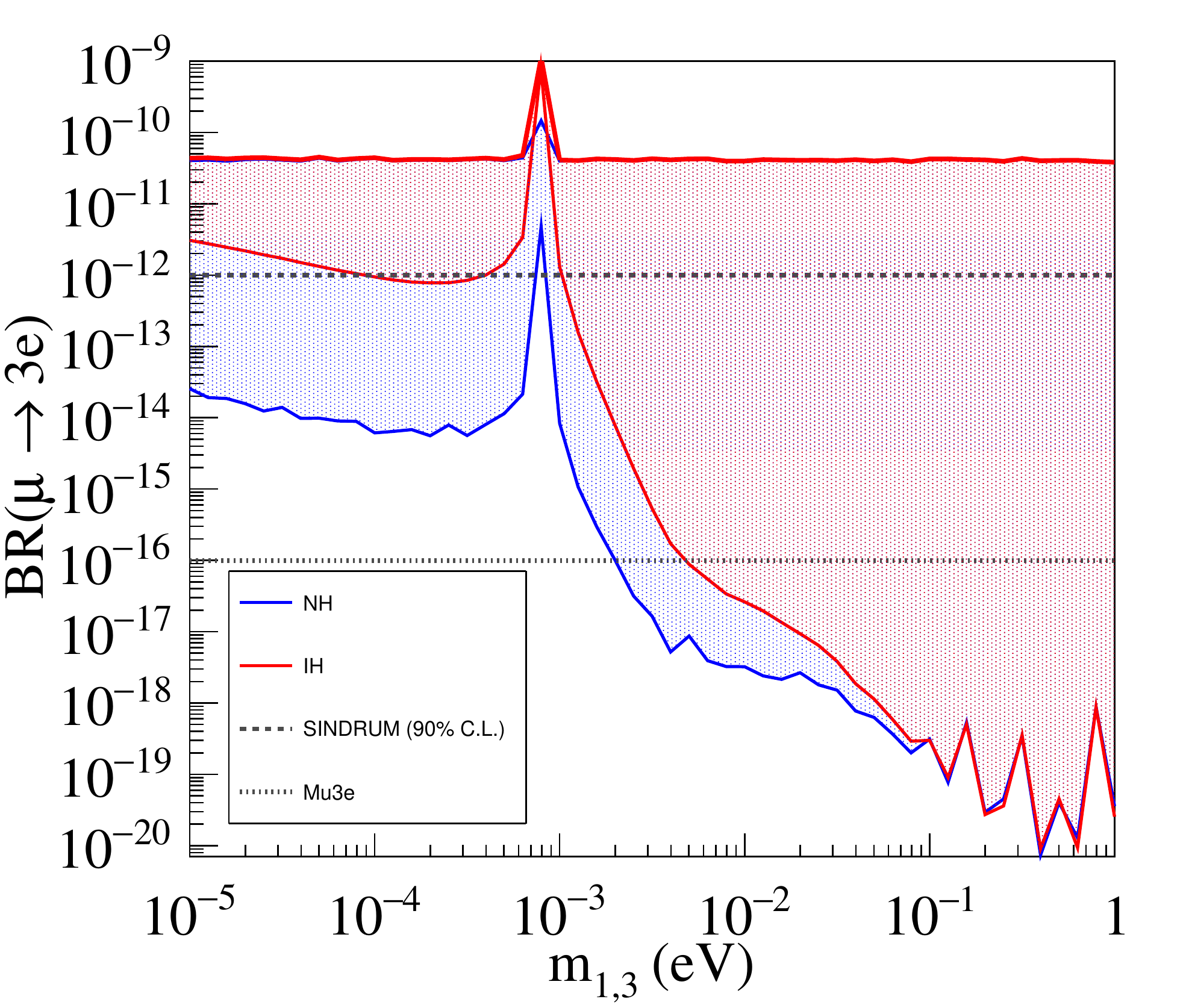}
	\caption{Branching ratio of the process $\mu \to 3e$ as a function of lightest neutrino mass, $m_1$ ($m_3$) for NH (IH) pattern.}
	\label{fig:muto3e}
\end{figure}

Fig.\ref{fig:muto3e} shows the variation of branching ratio of the process $\mu \to 3e$ with respect to the lightest neutrino mass where NH and IH pattern of light neutrno mass are represented by
blue and red bands respectively. The bounds on the branching ratio of this process are given by 
SINDRUM ($10^{-12}$) and Mu3e ($10^{-16}$) and the plot shows that both the bounds are satisfied by the model's predictions on this decay (due to light-heavy mixing) for both patterns of light neutrino mass.


Similarly, the bound on light neutrino mass from $\mue$ conversion rate with the mediation of heavy neutrino N and sterile neutrino S can be found as follows.
\begin{equation}
{\rm R}^{A(N,Z)}_{\mue} = \frac{\alpha_{\rm em}^3\alpha_W^4m_\mu^5}{16\pi^2\mwl^4\Gamma_{\rm capt}}
\frac{Z_{\rm eff}^4}{Z}\left|\mathcal{F}(-m_\mu^2)\right|^2 
\left(|\mathcal{Q}_L^W|^2+|\mathcal{Q}_R^W|^2\right), \label{eq:brmue_full}
\end{equation}
where the relevant parameters are given in appendix \ref{sec:mutoe}.

In Fig.\ref{fig:mutoecon} we have plotted the variation of $\mu \to e$ conversion against the lightest 
neutrino mass where the red curve represents the IH pattern of light neutrino mass and blue band represents the NH pattern of light neutrino mass. The experimental sensitivity of 
$\mu \to e$ conversion is represented by the horizontal dashed line. The plot shows that the model's predictions on decay rate of this conversion (due to contributions from heavy neutrino $N_R$ and sterile neutrino $S_L$) are not sensitive to the experimental bounds, for both NH and IH pattern of light neutrino mass. 

\begin{figure}[h!]
	\centering
	\includegraphics[width=0.49\textwidth]{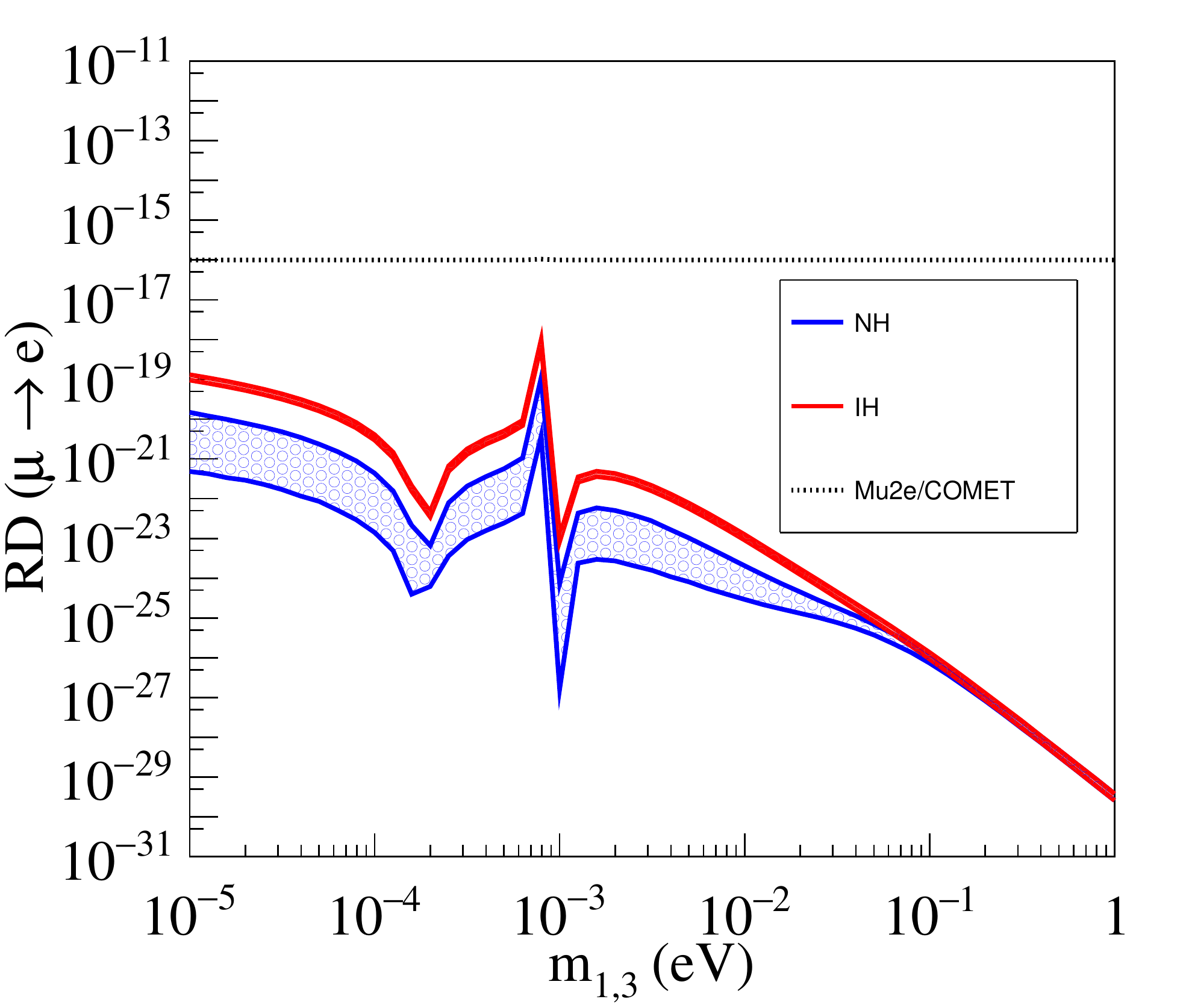}
	\caption{The rate for $\mu \to e$ conversion in Au nucleus as a function of the lightest neutrino mass, $m_1$ ($m_3$) for NH (IH) due to light-heavy mixing.}
	\label{fig:mutoecon}
\end{figure}
%

\section{Constraints from neutrinoless double beta decay}
\label{sec:0nubb}
We discuss here how the light-heavy neutrino mixing in left-right theories with type-II seesaw dominance leads to sizable new contributions to neutrinoless double beta decay. We give emphasis on left-handed current effects due to the exchange of heavy neutrinos $N_R$ and $S_L$. To ignore the effects of right-handed currents and the contributions of doubly charged scalar triplets to $0\nu\beta \beta$ transition we have assumed $M_{W_R} \gg M_{W_L}$ and large masses for scalar triplets. 
For a detailed discussion on various new physics contributions to neutrinoless double beta decay in TeV scale left-right symmetric model with large light-heavy neutrino mixing through type-II seesaw dominance one may refer ~\cite{Pritimita:2016fgr}.

\begin{figure}[htb!]
\centering
\includegraphics[scale=0.55]{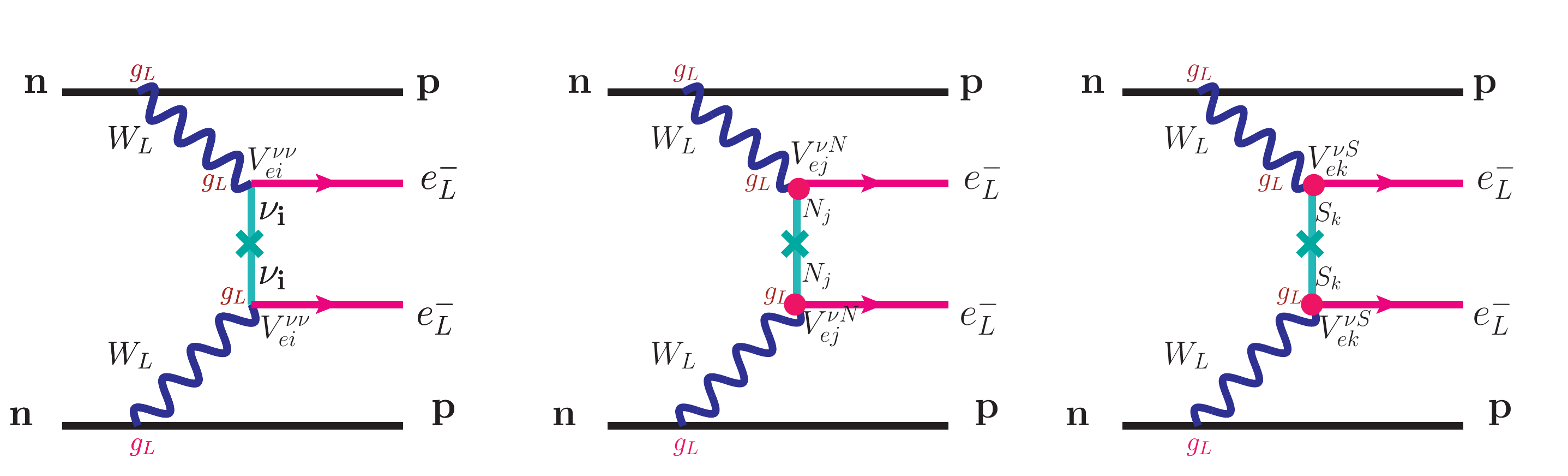}
 \caption{Feynman diagrams for $0\, \nu\, \beta \beta$ transition due to purely left-handed 
 currents with the exchange of virtual Majorana neutrinos $\nu_i$, $N_j$ and $S_k$.}
\label{feyn:lrsm-WLL}
\end{figure}
\subsection{Half-life and effective Majorana mass}
The relevant contributions arising from purely left-handed currents with the exchange of light active neutrinos (standard mechanism) and heavy neutrinos $N$ and $S$ are shown in Fig.\ref{feyn:lrsm-WLL}. The inverse half-life of $0\nu\beta \beta$ transition for a given isotope is,
\begin{eqnarray}
 \left[T_{1/2}^{0\nu}\right]^{-1} &=& G^{0\nu}_{01}\bigg[ |{\cal M}^{0\nu}_\nu \cdot \eta_\nu|^2 + 
 |{\cal M}^{0\nu}_N \cdot \left(\eta_{N}+ \eta_{S} \right) \big|^2 \bigg]
\label{eq:Hlife-a}
\end{eqnarray}
where the dimensionless lepton number violating parameters are given as,
\begin{eqnarray}
\label{eq:eta_LL} 
& &|\mathcal{\eta}_{\nu}| = \sum_{i=1,2,3} \frac{{\mbox{V}^{\nu \nu}_{ei}}^2\, m_{\nu_i}}{m_e} \,, 
  |\mathcal{\eta}_{N}| = m_p \sum_{i=1,2,3} \frac{{\mbox{V}^{\nu N}_{ei}}^2}{M_{N_i}} \,,                    
  |\mathcal{\eta}_{S}| = m_p \sum_{i=1,2,3} \frac{{\mbox{V}^{\nu S}_{ei}}^2}{M_{S_i}} \,.
\end{eqnarray}
In order to derive the bound on lightest neutrino mass by satuaring the current experimental bounds on half-life of a given isotope, one has to rewrite the inverse half-life in terms of a particle physics paremeter called Effective Majorana mass that contains lepton number violating information in it.
\begin{eqnarray}
 \left[T_{1/2}^{0\nu}\right]^{-1} &=&  G^{0\nu}_{01} \bigg| \frac{{\cal M}^{0\nu}_\nu}{m_e} \bigg|^2   \bigg[\big|m^{\nu}_{ee} \big|^2 
               +  \big|m^{N}_{ee} \big|^2+ \big|m^{S}_{ee} \big|^2 \bigg] \nonumber \\
               &=&G^{0\nu}_{01} \bigg( \frac{{\cal M}^{0\nu}_\nu}{m_e}\bigg)^2 \cdot |m^{\rm eff}_{\beta \beta}|^2\,.
\label{eq:Hlife-b}
\end{eqnarray}
Here $|m^{\rm eff}_{\beta \beta}|^2$ is the sum of contributions from light active neutrinos $\nu_L$, heavy right-handed neutrinos $N$ and sterile neutrinos $S$. Thus the half-life of neutrinoless double beta decay process is estimated by three kinds of contributions;
\begin{itemize}
\item Phase-space factor $G^{0\nu}$ which is responsible for detailed kinematics of the neutrinoless double beta decay process and is highly energy dependent,
\item Nuclear Matrix Elements (NMEs), $\mathcal{M}^{0\nu}_{\nu}$ and $\mathcal{M}^{0\nu}_{N}$ for light and heavy neutrinos that take care of the transition of the nucleus into daughter nuclei, 
\item Particle physics parameter called Effective Majorana mass $m^{\rm eff}_{\beta \beta}$ of the transition $2 d \rightarrow 2 u + 2 e^-$ inside the involved nucleons.
\end{itemize}
\begin{table}[h]
 \centering
\vspace{10pt}
 \begin{tabular}{lcccc}
 \hline \hline
 \\ \multirow{2}{*}{Isotope} & $G^{0\nu}_{01}$ $[{\rm yrs}^{-1}]$  
 & \multirow{2}{*}{${\cal M}^{0\nu}_\nu$} & \multirow{2}{*}{${\cal M}^{0\nu}_N$}  \\[2mm] 
\hline \\
$^{76}$Ge  & $5.77 \times 10^{-15}$  & 2.58--6.64 & 233--412  \\[1mm]
$^{136}$Xe  & $3.56 \times 10^{-14}$ & 1.57--3.85 & 164--172  \\[1mm] 
\hline \hline
 \end{tabular}
 \caption{$G^{0\nu}_{01}$ and NMEs~\cite{Meroni:2012qf}}
 \label{tab:nucl-matrix}
\end{table} 
The values of $G^{0\nu}$ and NMEs are different for different isotopes as presented in Table.\ref{tab:nucl-matrix} and the effective Majorana mass parameter $m^{\rm eff}_{\beta \beta}$ is expressed in terms half-life of a given isotope as 
\begin{eqnarray}
 &&\left[T_{1/2}^{0\nu}\right]^{-1} = G^{0\nu}_{01} |\mathcal{M}^{0\nu} (A) |^2 \cdot \bigg(\frac{m^{\rm eff}_{\beta \beta}}{m_e} \bigg)^2 \, \nonumber \\
 &&m^{\rm eff}_{\beta \beta} = \frac{m_e}{\sqrt{G^{0\nu}_{01} T^{0\nu}_{1/2}}}\,.
\label{eq:Hlife-c}
\end{eqnarray}
Using Table.\ref{tab:nucl-matrix} and Table.\ref{tab:half-life}, one can derive theoretical limits on effective Majorana mass which we use in our numerical estimations to derive limits on absolute value of light neutrino masees.

\begin{table}[htp]
 \centering
 \caption{Limits on the half-life of $0\nu\beta\beta$.}
 \label{table:halflifelimits}
 \vspace{7pt}
 \begin{tabular}{lc}
  \hline \hline 
   Experiment & Limit \\[1mm]
  \hline HM & $1.9 \times 10^{25}\ {\rm yrs}$ \\
  GERDA & $2.1\times 10^{25}\ {\rm yrs}$ \\
  Combined $^{76}$Ge & $3.0\times 10^{25}\ {\rm yrs}$ \\
  GERDA Phase II & $5.2\times 10^{25}\ {\rm yrs}$ \\
  \hline EXO & $1.6\times 10^{25}\ {\rm yrs}$ \\
  KamLAND-Zen & $1.9\times 10^{25}\ {\rm yrs}$ \\
  Combined $^{136}$Xe & $3.4\times 10^{25}\ {\rm yrs}$ \\
  nEXO & $6.6\times 10^{27}\ {\rm yrs}$ \\
  \hline \hline
 \end{tabular}
  \label{tab:half-life}
\end{table}

\subsection{Bound on lightest neutrino mass from $0\nu\beta\beta$}
\noindent
{\bf Standard mechanism: $m^{\nu}_{ee}$ from light neutrinos $\nu$ :-}
 The standard contribution to effective Majorana mass due to exchange of light neutrinos is given by,
$$m^{\nu}_{ee} = \sum_{i=1,2,3} \left(\mbox{V}^{\nu \nu}\right)^2_{ei}\, m_{i}\, m_{i} \simeq \sum_{i = 1}^3 (U_\text{PMNS})_{e i}^2 \, m_i\,.$$
where $V^{\nu\nu}$ is the mixing matrix containing non-unitarity effects which is approximated to be $U_{\rm PMNS}$.

\begin{figure}[h!]
\centering
\includegraphics[width=0.49\textwidth]{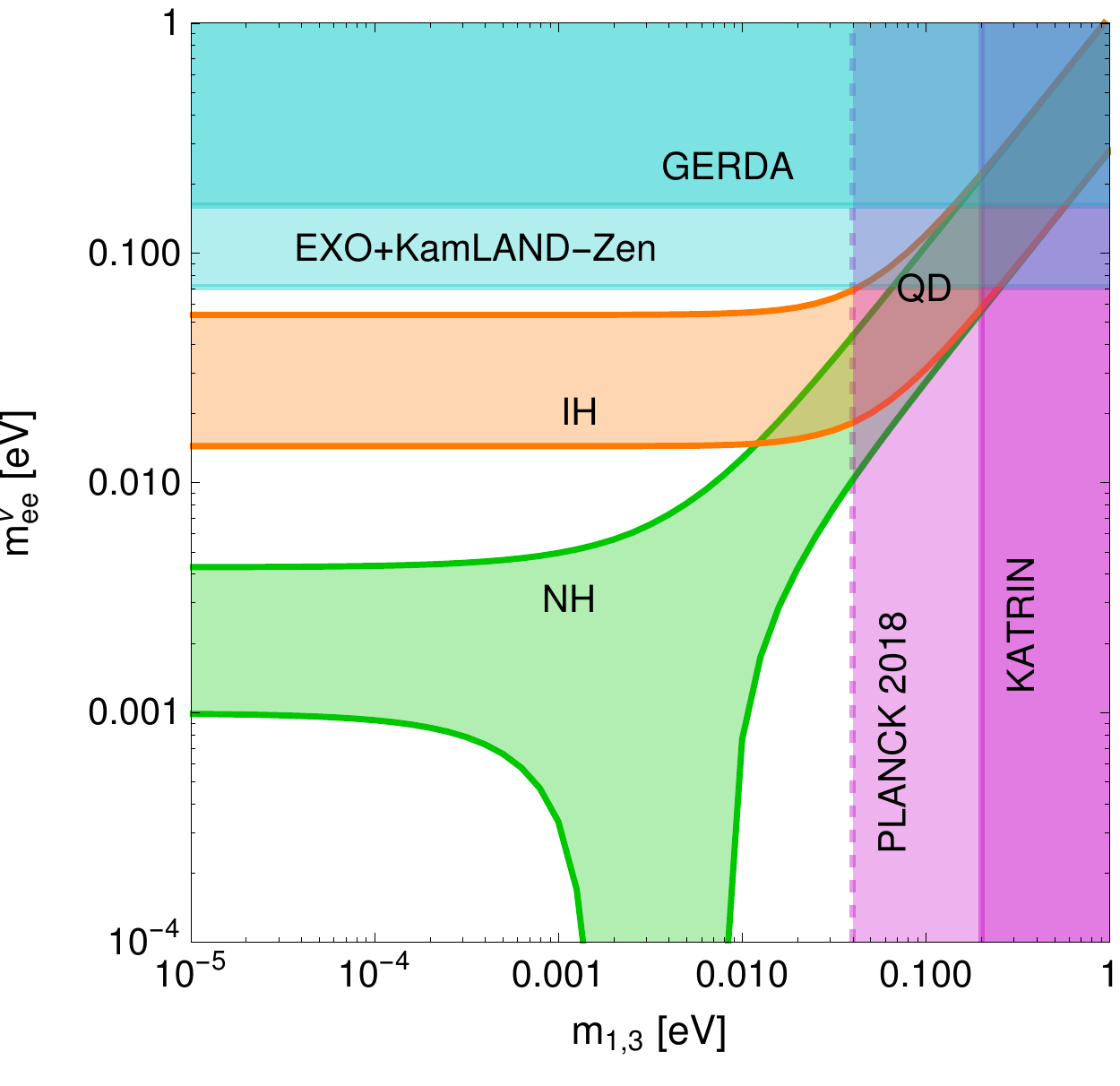}
\includegraphics[width=0.49\textwidth]{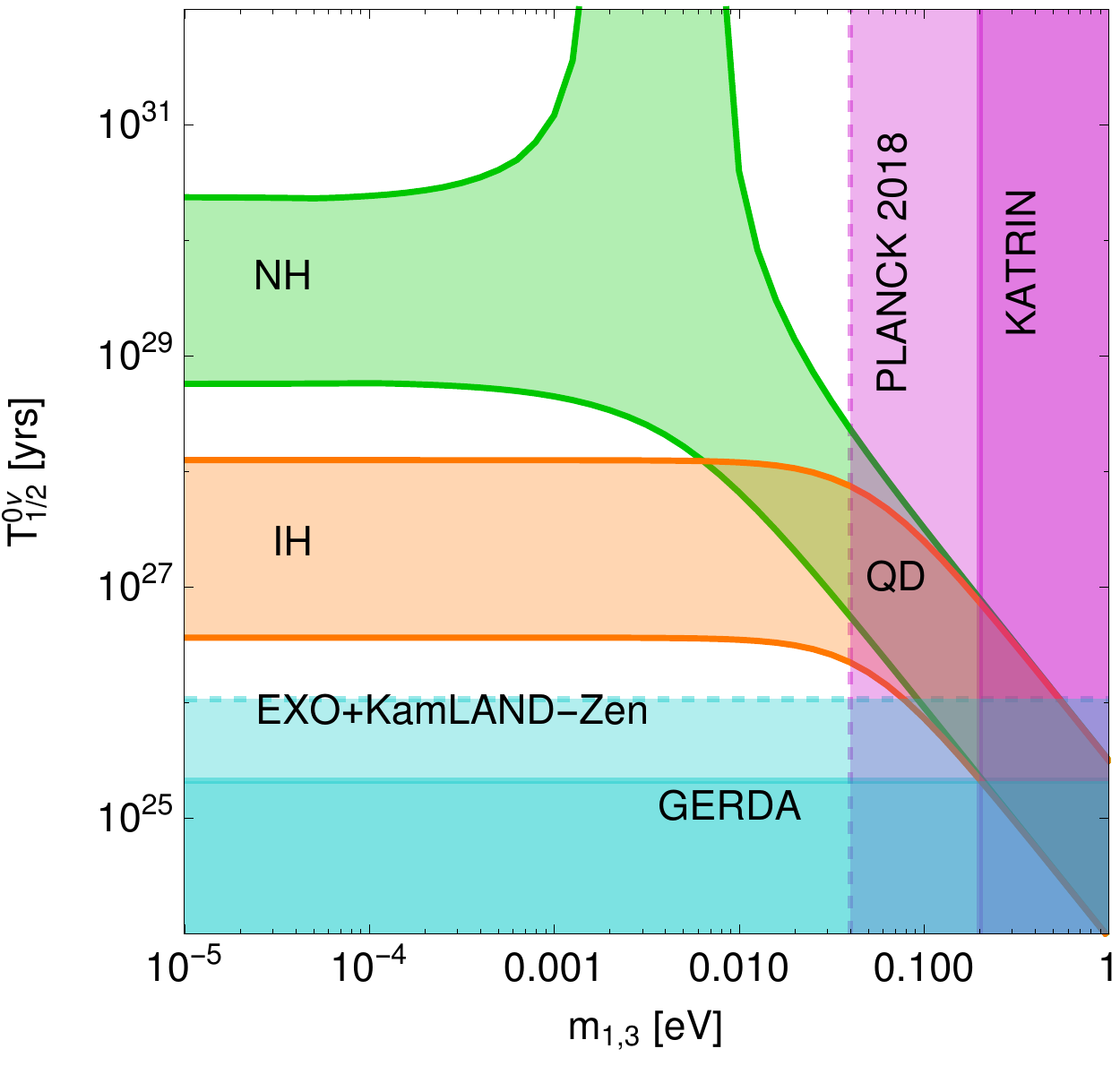}
\caption{Left panel: Effective Majorana mass parameter ($|m_{ee}|$) due to standard mechanism as a function of lightest neutrino mass $m_1$ ($m_3$) for NH (IH). Right panel: SM contribution to the half-life of $0\nu\beta\beta$ transition as a function of lightest neutrino mass. The blue horizontal bands show the limits on effective Majorana mass and half-life from GERDA and EXO+KamLAND-Zen experiments. The vertical purple bands are for the constraints on sum of light neutrino masses from cosmology data (PLANCK 2018) and KATRIN detector.}
\label{fig:mee-std}
\end{figure}

\begin{figure}[h!]
\centering
\includegraphics[width=0.49\textwidth]{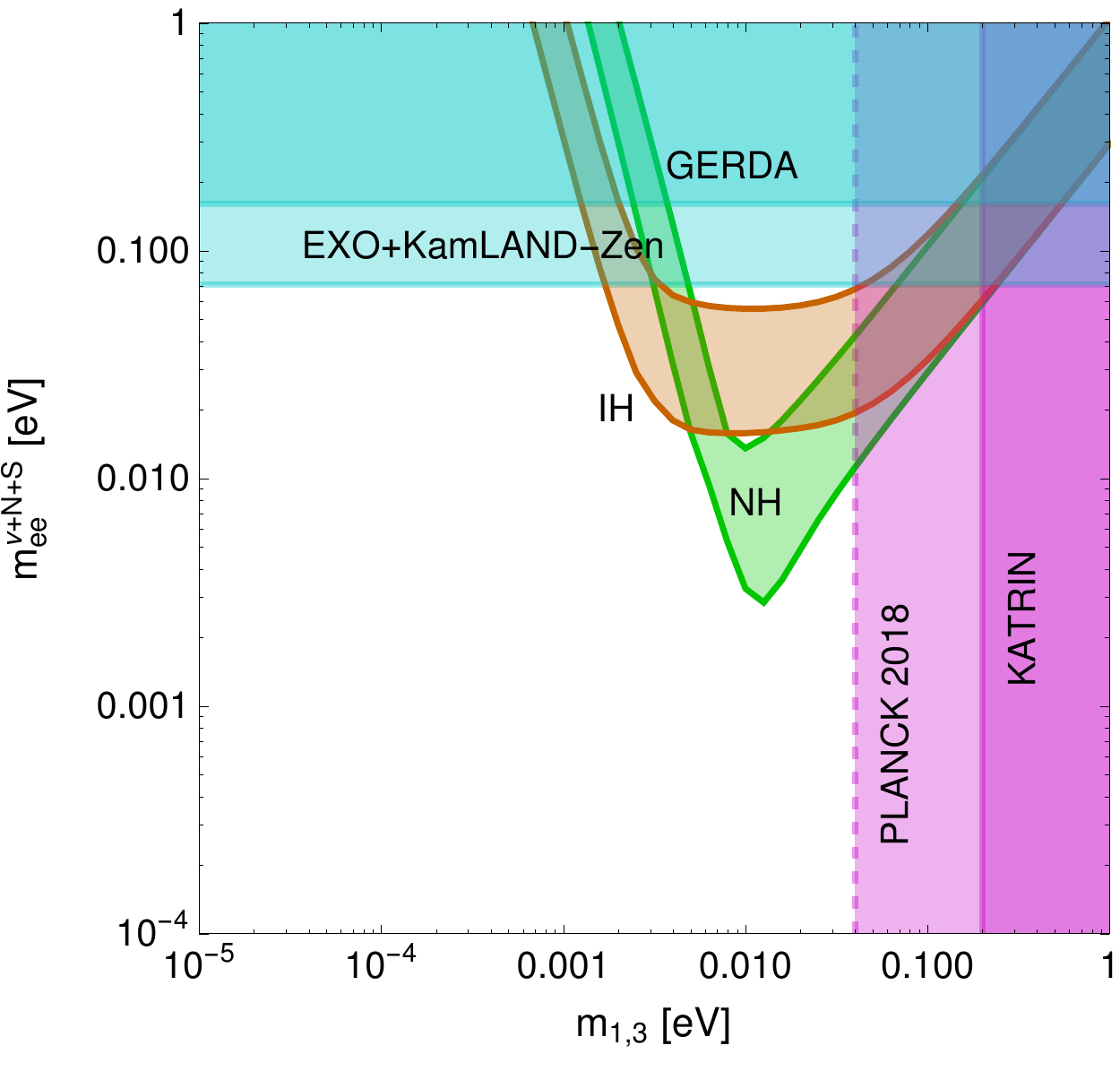}
\includegraphics[width=0.49\textwidth]{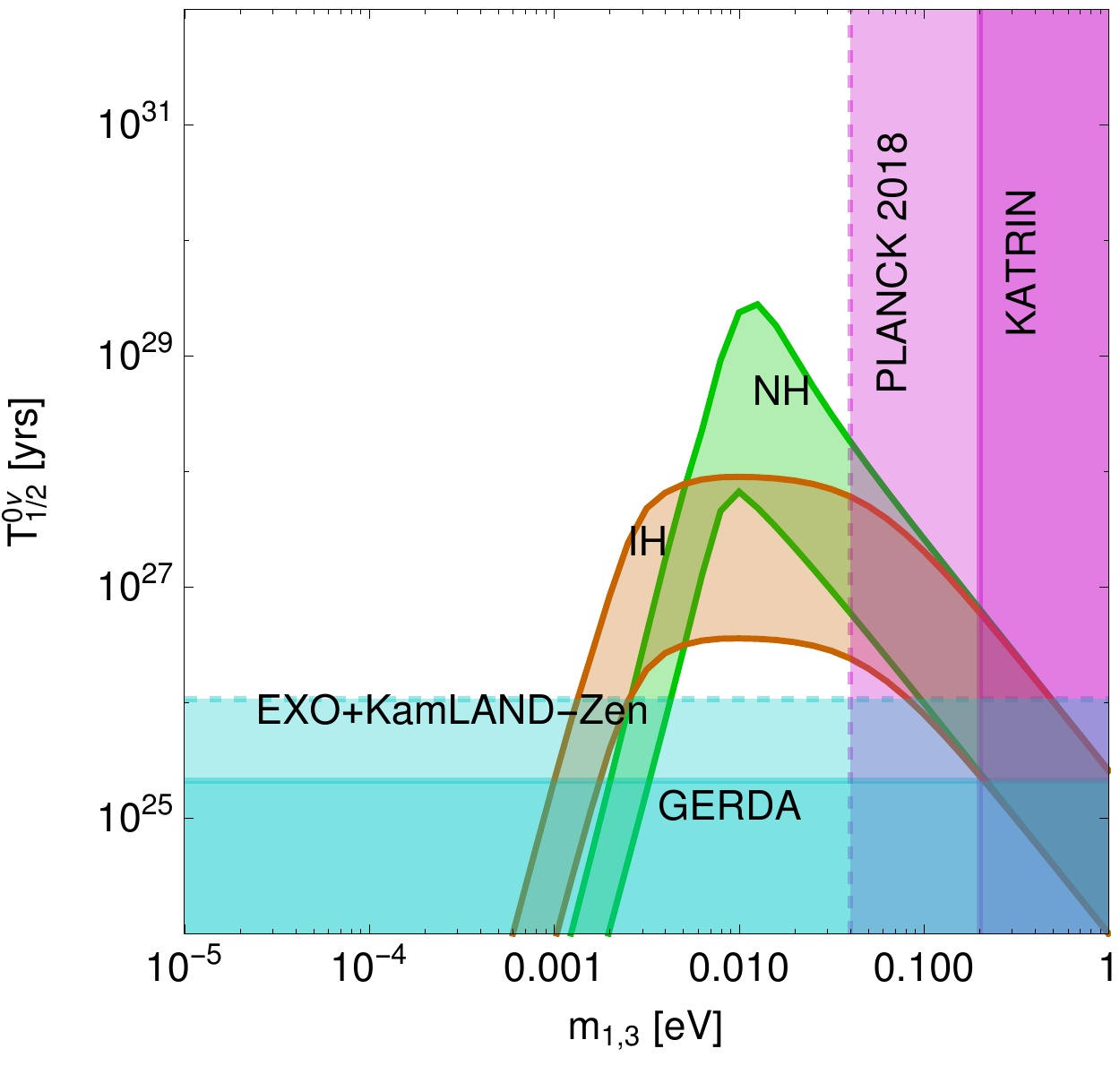}
\caption{Effective Majorana mass (left panel) and half-life (right panel) as a function of lightest neutrino mass $m_1$ ($m_3$) for NH (IH) due to the combined contributions of standard mechanism, $N_R$ and $S_L$ mediated diagrams.}
\label{fig:0nubb-total}
\end{figure}

\noindent
{\bf New physics contribution: $m^{N,S}_{ee}$ from heavy neutrinos $N_R$ and $S_L$ :-}
\begin{figure}[h!]
\centering
\includegraphics[width=0.49\textwidth]{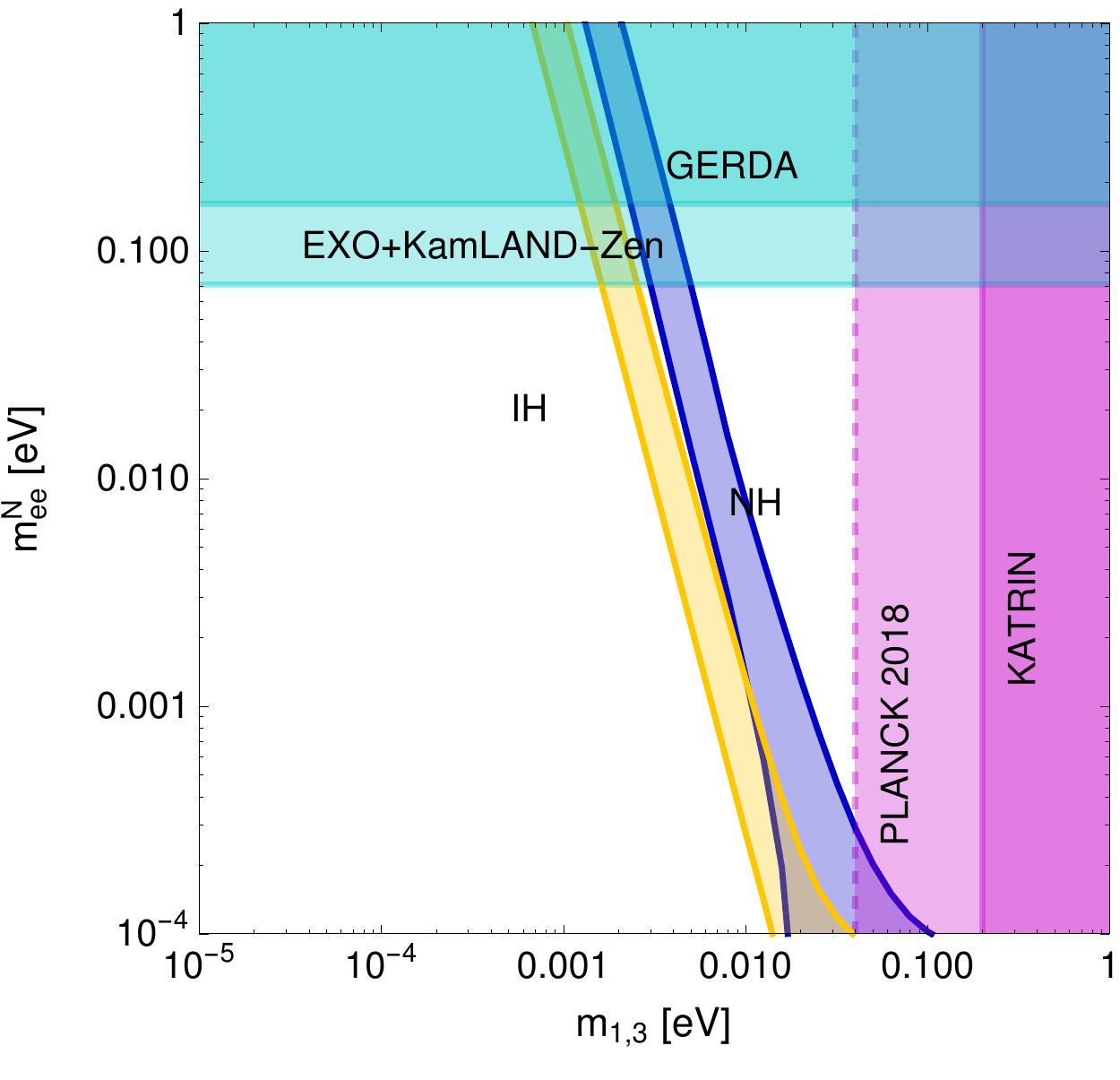}
\includegraphics[width=0.49\textwidth]{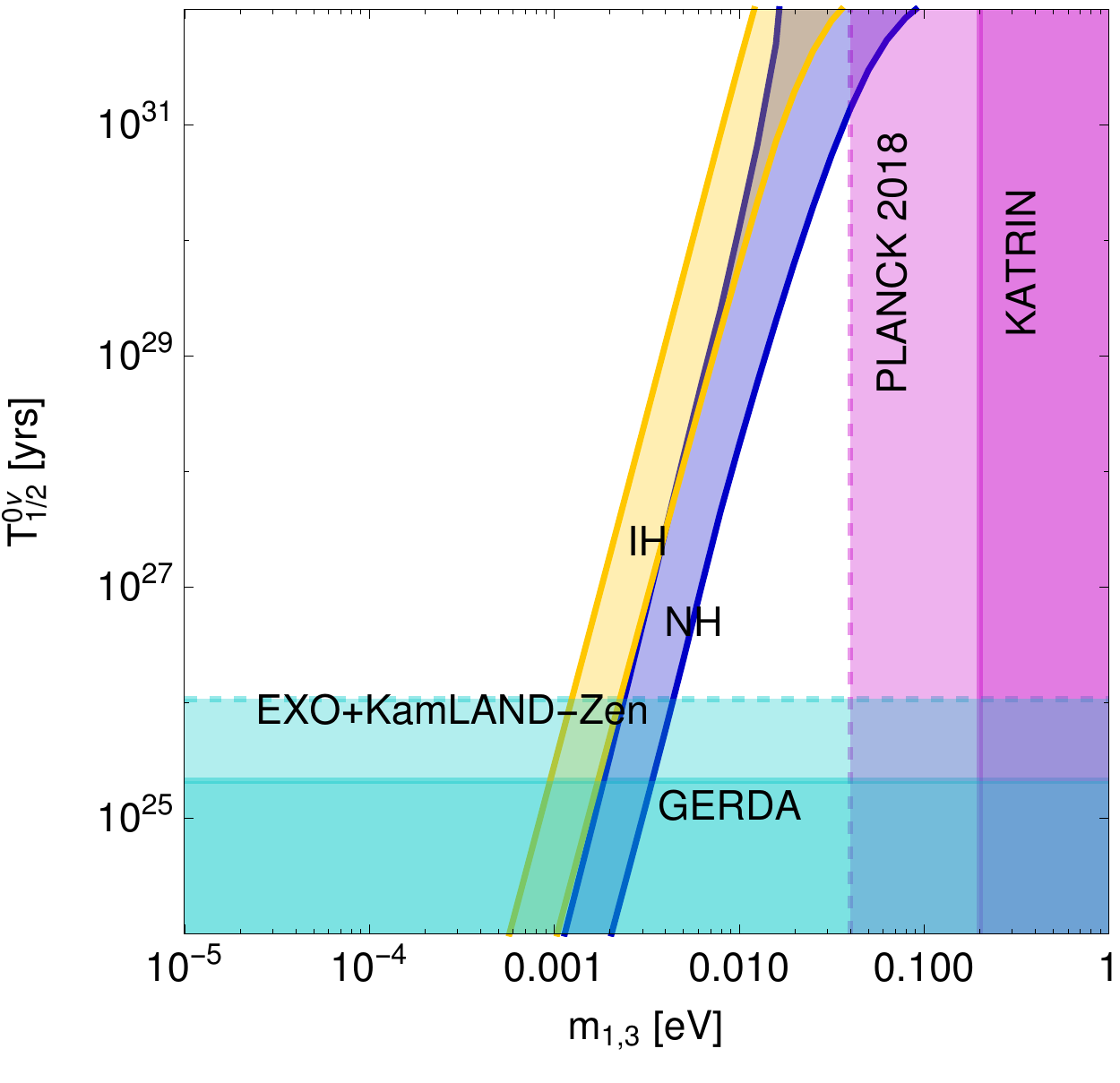}
\caption{Left panel: New physics contribution to the plot of effective Majorana mass as a function of lightest neutrino mass, $m_1$ ($m_3$) for NH (IH) via $W_L-W_L$ channel with the exchange of heavy neutrino $N_R$ and sterile neutrino $S_L$. Right panel: Contributions of $N_R$ and $S_L$ to the plot of half-life vs lightest neutrino mass.}
\label{fig:0nubb-NS}
\end{figure}
The effective Majorana masses due to the contributions of heavy neutrinos $N_R$ and $S_L$ to $0\nu\beta\beta$ decay (represented by the second and third Feynman diagrams in Fig.\ref{feyn:lrsm-WLL}) are as follows,
\begin{eqnarray}
 &&|m^{N}_{ee} | = \langle p^2 \rangle \sum_{i=1,2,3} \frac{\left(\mbox{V}^{\nu N}\right)^2_{ei}}{M_{N_i}} \,, \nonumber \\
 &&|m^{S}_{ee} | = \langle p^2 \rangle \sum_{i=1,2,3} \frac{\left(\mbox{V}^{\nu S}\right)^2_{ei}}{M_{S_i}} \,.
\end{eqnarray}
Here $M_{N_i} (M_{S_i})$ are the mass eigenvalues of heavy right-handed (sterile) neutrinos. It is to be noted here that the type-II seesaw dominance scheme in our model establishes a relationship between these mass eigenvalues and the light neutrino masses. $\mbox{V}^{\nu N}$ and $\mbox{V}^{\nu S}$ are the mixing matrices which represent the mixing between light-heavy neutrinos ( $\nu$ and $N_R$) and active-sterile neutrinos ($\nu$ and $S_L$) respectively. The neutrino virtual momentum $\langle p^2 \rangle$ plays an important role as the half-life formula for $0\nu\beta \beta$ transition differs for $m_i(M_i) \ll \langle p^2 \rangle$ or $m_i(M_i) \gg \langle p^2 \rangle$. The typical expression for neutrino virtual momentum $\langle p^2 \rangle$ is written as,
\begin{eqnarray}
\langle p^2 \rangle = - m_e\,m_p\, \frac{{\cal M}^{0\nu}_N}{{\cal M}^{0\nu}_\nu} \simeq (\mbox{200\, MeV})^2\,.
\end{eqnarray}
Fig.\ref{fig:mee-std} shows the effective Majorana mass parameter in the left panel and half-life in the right panel due to the standard mechanism as a function of lightest neutrino mass $m_1$(NH) and $m_3$(IH). Both the plots show that the quasi-degenerate (QD) pattern of light neutrinos, i.e. $m_1\simeq m_2\simeq m_3$ is disfavoured by cosmology data on sum of light neutrino masses whereas the NH(green band) and IH(red band) patterns may not be probed even by the next generation experiments. This propels the idea of exploring possible new physics contributions to neutrinoless double beta decay which might give a hint on mass hierarchy and lightest neutrino mass. Such possibility occurs when the contributions of heavy neutrino $N_R$ and sterile neutrino $S_L$ are considered. The same is shown in Fig.\ref{fig:0nubb-NS}. The contributions of $N_R$ and $S_L$ to effective Majorana mass parameter and half-life predictions makes both NH(blue band) and IH(orange band) patterns of light neutrino masses sensitive to the current experimental bounds. In Fig.\ref{fig:0nubb-total} the contributions of all three types of neutrinos ($\nu, N_R, S_L$) are summed up and plotted against lightest neutrino mass. In this case also the both NH(green band) and IH(red band) patterns saturate the current experimental bounds on effective Majorana mass and half-life. In all the three figures the horizontal blue bands stand for the improved limits on $0\nu\beta\beta$ decay on effective Majorana mass and half-life with the combined results of GERDA and KamLAND-Zen experiments. The vertial magenta region is disfavoured by Planck-2018~\cite{Tian:2020tur} and KATRIN data~\cite{Aker:2019uuj} on sum of light neutrino masses. In these plots the largest value of right-handed neutrino mass is fixed at $M_{N_R}=1$ TeV while keeping $M_{W_R}, M_{\Delta^{++}} >> M_{N_R}$. The Majorana phases and Dirac CP-phase are varied between $0 \to \pi$ for all the plots while other neutrino oscillation parameters are taken in their allowed $3\sigma$ range.

 The predictions of our model on lightest neutrino mass due to heavy and sterile neutrino contributions to neutrinoless double beta decay are as follows,
\begin{itemize}
 \item When the contributions of only $N_R$ and $S_L$ (excluding standard contribution) are considered, the allowed values of lightest neutrino mass are found to be in the range of $10-25$~meV for $m_1$(NH) and $25-40$~meV for $m_3$(IH). This is done by saturating the effective Majorana mass $m^{N+S}_{ee}$ with the current GERDA and KamLAND-Zen bounds. It remains the same for half-life.
 \item When the contributions of all three types of neutrinos ($\nu, N_R, S_L$) are considered the predicted values of lightest neutrino mass lies in the range of $2-15$~meV for $m_1$(NH) and $15-25$~meV for $m_3$(IH) by sturating $m^{\nu+N+S}_{ee}$ with the current experimental limits. 
\end{itemize}
Using these predicted values of lightest neutrino mass other light neutrino mass eigenvalues and heavy neutrino masses can be derived.

\section{Complementary studies between LNV and LFV Decays}
\label{sec:lnv-lfv}
\subsection{Correlation between $m_{\beta}$ and sum of light neutrino masses $\sum m_i$}
\begin{figure}[htb!]
\centering
\includegraphics[width=0.49\textwidth]{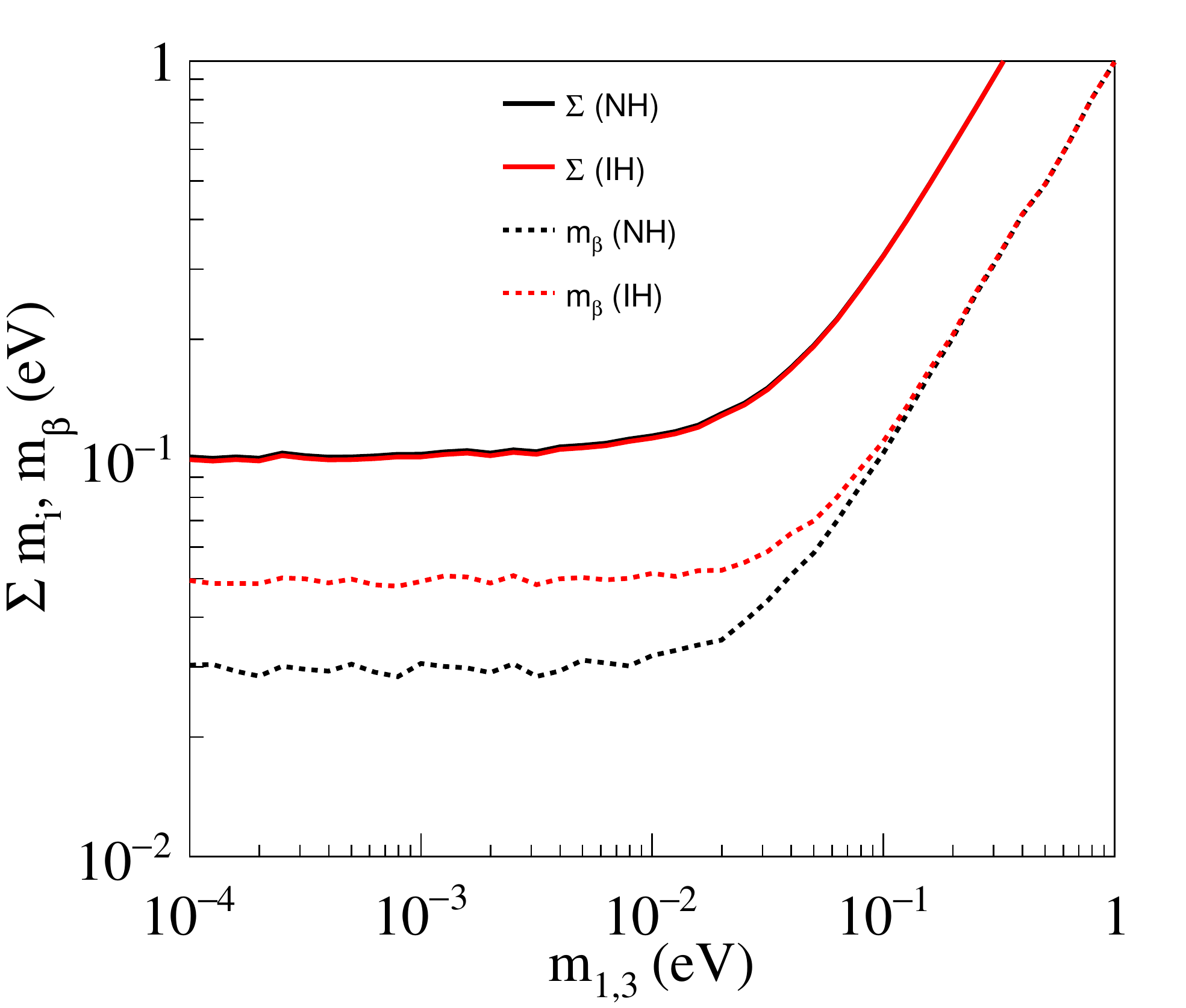}
\caption{Plot showing the variation of $\sum m_i$ and $m_\beta$ with lightest neutrino mass ($m_1$ for NH, $m_3$ for IH).}
\label{fig:mbetamsum}
\end{figure}

In this subsection we examine how the combined constraints from single beta decay and cosmology can limit absolute scale of light neutrino mass. The important parameter for single $\beta$ decay sensitive to the electron neutrino mass is defined as follows.
\begin{equation}
 m_{\beta} = \sqrt{\sum_{i} \mid U^2_{ei} m^2_i\mid}
\end{equation}
In the present model $ U^2_{ei}$ is replaced by $\mathcal{V}^{\nu \nu^2}_{ei}$ due to sizeable non-unitarity effects. The current limit on $m_{\beta}$ from KATRIN experiment is $m_{\beta} < 1.1\,$eV while an improved bound of $0.2$\,eV~\cite{Aker:2019uuj} is expected in the future. The sum of light neutrino masses is tightly constrained by Planck 2018 data~\cite{Tian:2020tur} which is $\sum m_i=m_1+m_2+m_3 < 0.12\,$~eV. Fig.\ref{fig:mbetamsum} shows the variation of $\sum m_i$ and $m_{\beta}$ with lightest neutrino mass.

\subsection{Correlation between $m_{ee}$ and sum of light neutrino masses $\sum m_i$}
\begin{figure}[h!]
\centering
\includegraphics[width=0.49\textwidth]{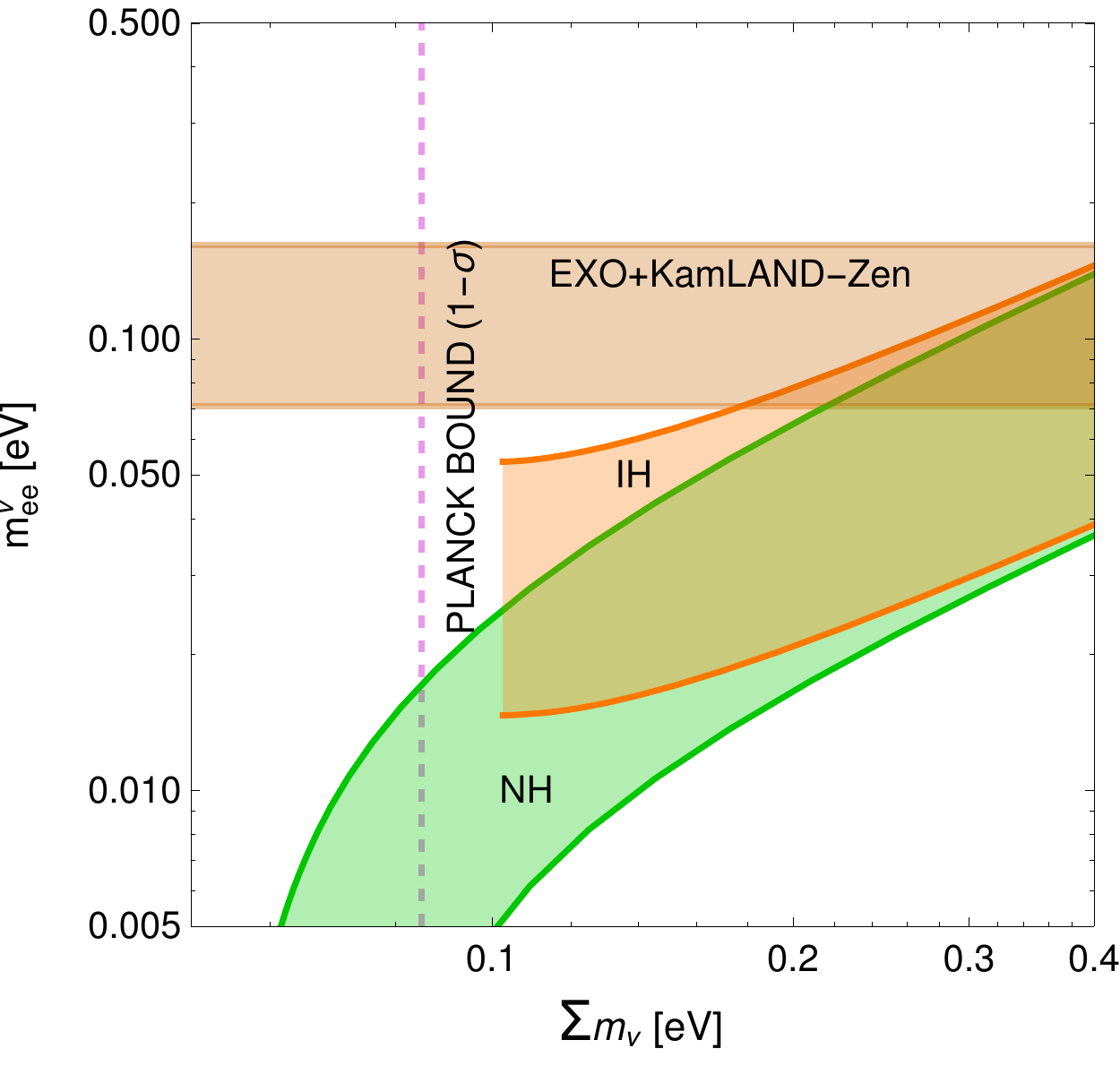}
\includegraphics[width=0.49\textwidth]{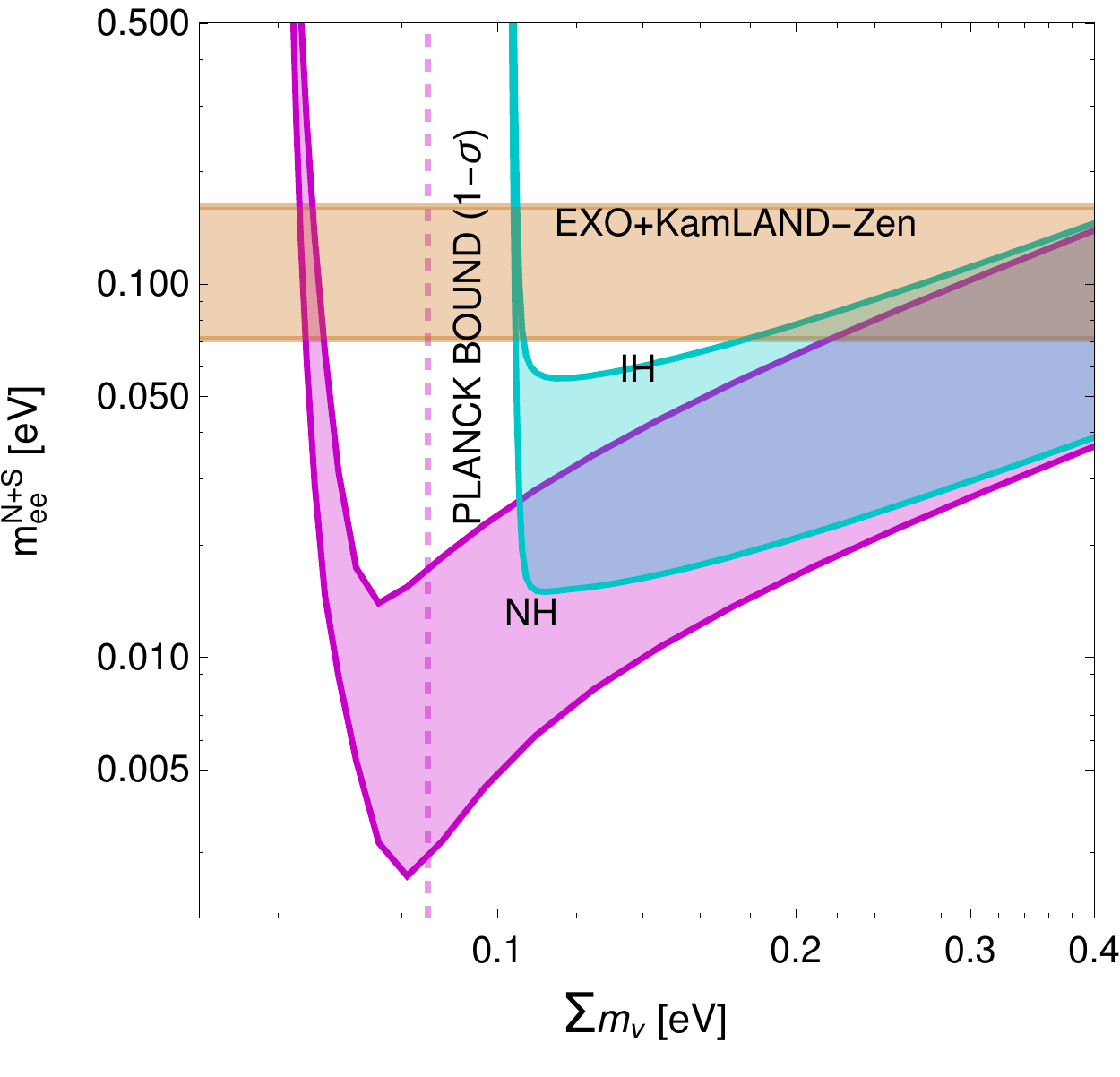}
\caption{Allowed region of effective Majorana mass parameter ($|m_{ee}|$) as a function of sum of light neutrino masses ($\Sigma m_i$) for standard mechanism (left-panel) and new physics contributions from $N_R$ and $S_L$ mediated diagrams (right-panel).}
\label{fig:mee-summass}
\end{figure}

Here we discuss the model predictions on effective Majorana mass by changing the sum of light neutrino masses $\Sigma m_i$. We use the following limits on sum of light neutrino masses~\cite{Seljak:2004xh,Costanzi:2014tna,Palanque-Delabrouille:2014jca} for displaying 
the allowed region of $m_{ee}$ in Fig.\ref{fig:mee-summass}.
\begin{eqnarray}
&&m_\Sigma < \mbox{84\,meV}\quad \quad ~(1\sigma~\mbox{C.L.}) \nonumber \\
&&m_\Sigma < \mbox{146\,meV}\quad \quad (2\sigma~\mbox{C.L.}) \nonumber \\
&&m_\Sigma < \mbox{208\,meV}\quad \quad (3\sigma~\mbox{C.L.}) 
\end{eqnarray}
In both the plots of Fig.\ref{fig:mee-summass} the horizontal bands represents the experimental bounds on effective Majorana mass $m_{ee}$ by GERDA and EXO+KamLAND-Zen while the vertical dashed line shows the bound on sum of light neutrino masses from cosmology. The plot in the left-panel shows the variation of effective Majorana mass parameter due to standard mechanism with sum of light neutrino masses. It shows that both NH (green band) and IH (red band) patterns of light neutrino masses are not sensitive to the current experimetal bounds on $m_{ee}$ and also disfavoured by cosmology when only  light neutrino ($\nu$) contribution is considered. The plot in the right-panel shows the variation of effective Majorana mass parameter due to new physics contributions (contributions of $N_R$ and $S_L$) with sum of light neutrino masses. It shows that NH (purple band) pattern of light neutrino masses saturates the experimental bound on $m_{ee}$ as well as the cosmology bound on sum of light neutrino masses whereas IH (blue band) pattern is disfavoured by cosmology. Thus the model with type-II seesaw dominance gives an important result on the hierarchy of light neutrino masses when the contributions of heavy and sterile neutrinos are considered.

We also do a comparative analysis between neutrinoless double beta decay and LFV processes to examine how light-heavy neutrino mixing is constrained from a combined study of LNV and LFV.

\begin{figure}[htb!]
\centering
\includegraphics[width=0.49\textwidth]{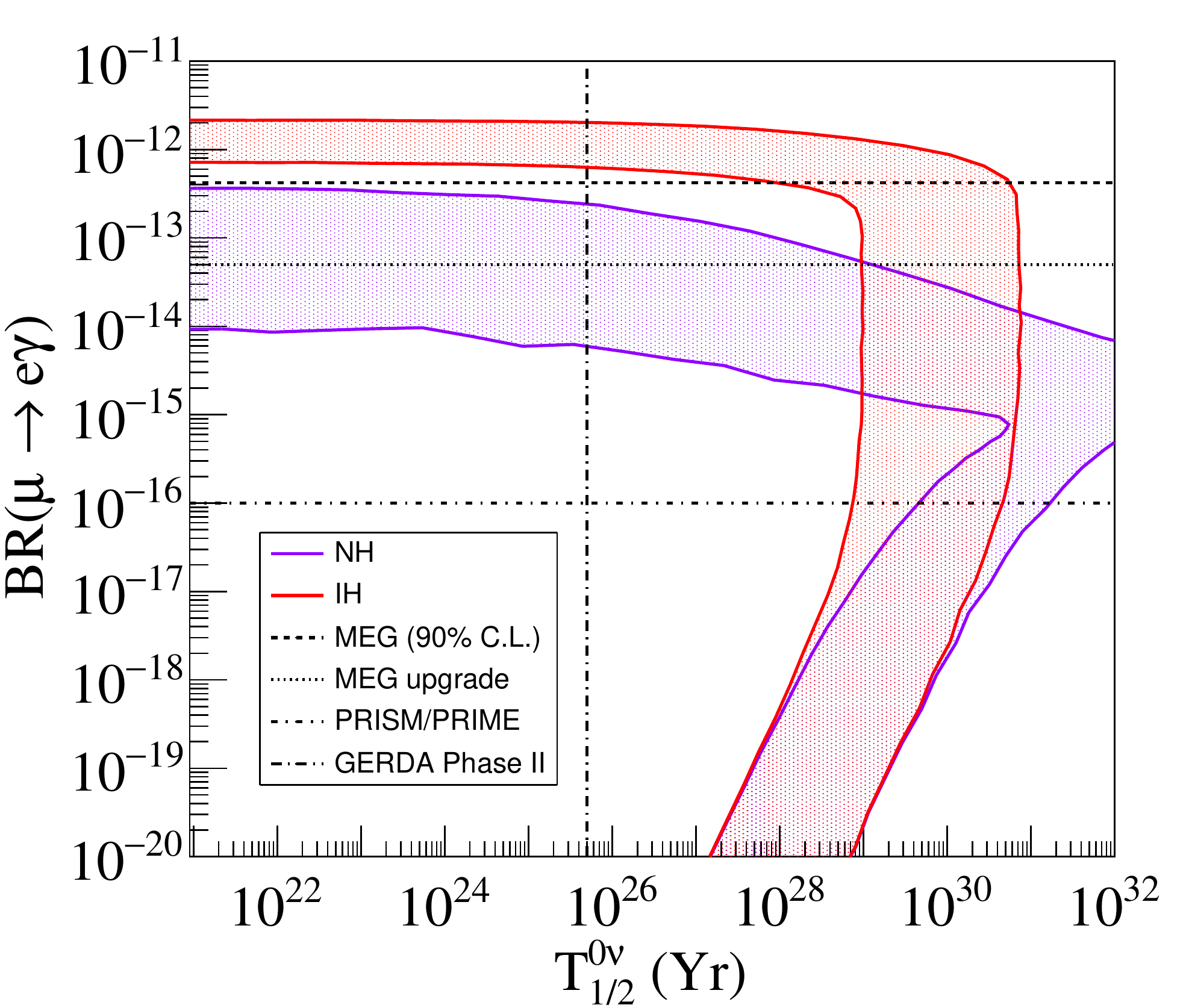}
\caption{Correlation plot between LFV processes $\mu \to e \gamma$ and LNV $0\nu\beta\beta$ 
decay for $^{76}$Ge isotope.}
\label{fig:lnvlfv}
\end{figure}

\subsection{Electric dipole moment (EDM) of charged leptons}     
The dipole moment of electron is an important process whose experimental observation 
will reveal violation of Parity $P$ and Time reversal $T$ symmetry (or violation of 
CP invariance) in nature. The electric dipole moment formula derived for charged 
leptons $\ell_\alpha (\alpha=e, \mu, \tau)$ is given by~\cite{Nieves:1986uk,Nemevsek:2012iq,
	Tello:2012qda,Barry:2013xxa}
\begin{eqnarray}
d_\alpha = \frac{e \alpha_{\rm \small W}}{8 \pi M^2_{W_L}} 
\mbox{\Large Im}\bigg[\sum^{3}_{i=1} {\mbox{V}}^{\nu N}_{\alpha i} {\mbox{V}}^{N N}_{i \alpha}\, \xi \,
\mathcal G^{\gamma}_{2}(x_{N_i}) M_{N_i}\bigg]\,,
\end{eqnarray}
where $\alpha_W \simeq 1/30$ is weak fine structure constant, $M_{W_L}=80.3$~GeV, $\xi \leq 10^{-4}$ 
is the $W_L-W_R$ mixing and $M_N$ is the mass matrix for heavy neutrinos. The mixing matrices are 
$\mbox{V}^{\nu N}=(v_L/v_R) M_D U_{\rm PMNS} m^{-1}_\nu$, $\mbox{V}^{N N} =U_{\rm PMNS}$. 
The resulting dipole moment for electron can be expressed as a function of lightest neutrino mass and one can derive similar bound on absolute scale lightest neutrino mass by saturating the experimental bound. 

\begin{figure}[h]
	\centering
	\includegraphics[width=0.59\textwidth]{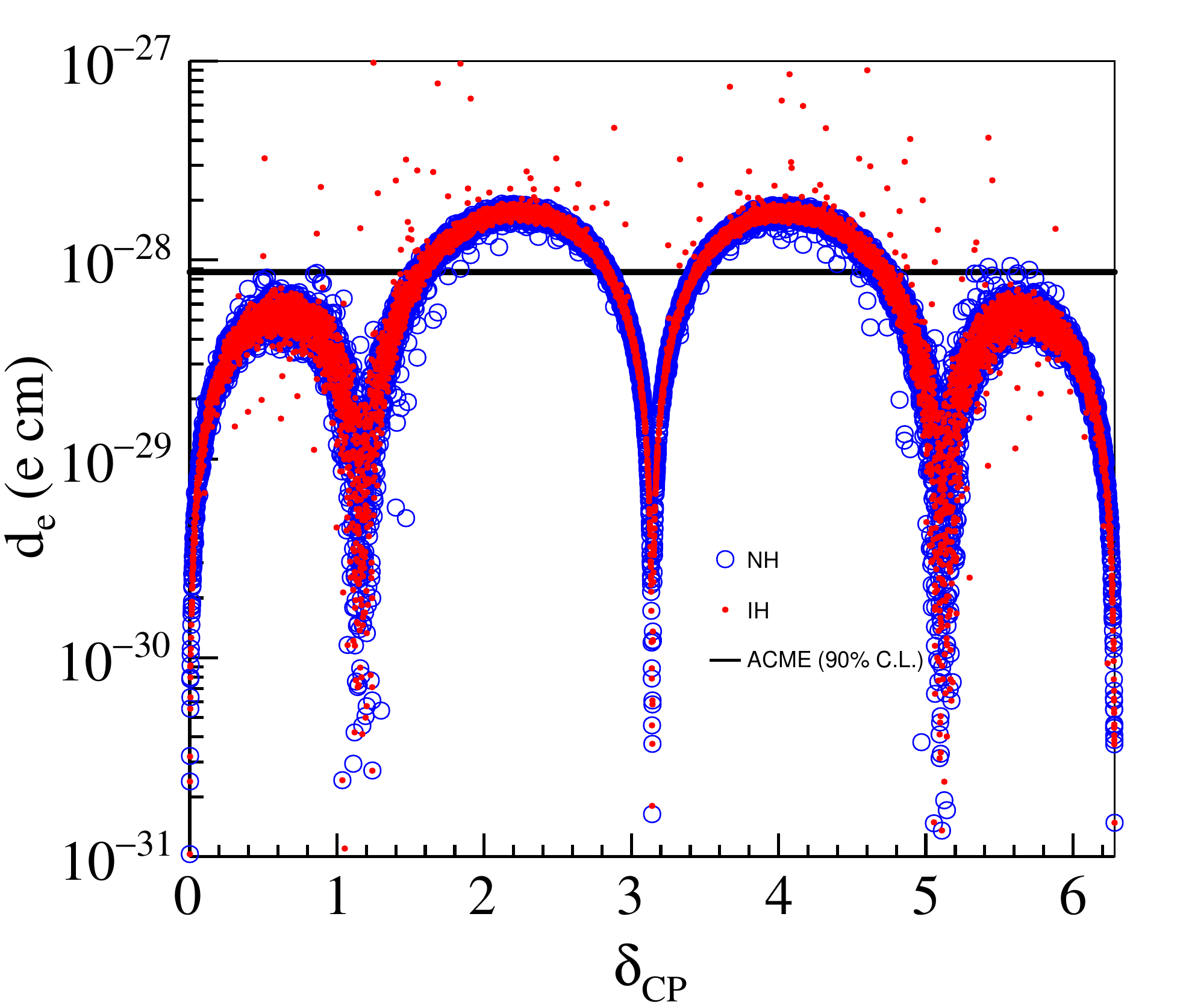}
	\caption{Electric dipole moment, $d_e$ as a function of Dirac CP-phase, $\delta_{\rm CP}$ for NH and IH patterns of light neutrino masses considering diagonal structures for Dirac neutrino mass matrix $M_D$ and $N_R - S_L$ mixing matrix $M$.}
	\label{fig:edm_deltaCP}
\end{figure}

In the present work, we can go to a basis where both Dirac neutrino mass matrix 
$M_D$ connecting $\nu_L-N_R$ and heavy Dirac mass $M$ connecting $N_R-S_L$ can be diagonal simultaneously. Taking $M_D \simeq 80 \mbox{diag}(m_e, m_\mu, m_\tau)$~GeV 
and $M_D \simeq 500 \mbox{diag}(1, 1, 1)$~GeV, we found that the dipole moment only depends 
upon PMNS phases contributing to the imaginary part. Considering vanishing Majorana phases, the variation of electron dipole moment with Dirac CP-phase $\delta_{\rm CP}$ is 
shown in Fig.\ref{fig:edm_deltaCP} The horizontal line represents the current bound on electron EDM set by ACME \cite{Baron:2013eja}.

\section{Comments on muon (g-2) anomaly}
\label{sec:muong2}
An intensive activity is going on to explain the recent results declared by Fermi National Accelerator Laboratory 
(FNAL) on the measurement of muon magnetic dipole moment, $a_{\mu}= \frac{g_{\mu}-2}{2}$ that just confirmed the existing anomaly in the muon sector. The theoretical value of $a_{\mu}$ estimated by Standard Model \cite{Tanabashi:2018oca,Blum:2013xva,Bennett:2006fi,Abi:2021gix}, the measured value by Brookhaven National Laboratoty (BNL) \cite{Bennett:2006fi} and the recent result by FNAL \cite{Abi:2021gix} are as follows.
\begin{eqnarray}
a_{\mu}^{\text{SM}} = (11659181.0 \pm 4.3) \times 10^{-10}   \\
a_{\mu}^{\text{BNL}} = (11659208.0 \pm 6.3)\times 10^{-10}  \\
a_{\mu}^{\text{FNAL}} = (11659204.0 \pm 5.4)\times 10^{-10}
\end{eqnarray}
Combining both the results of BNL and FNAL, the new world average now stands at,
\begin{equation}
a_{\mu}^{\text{2021}} = (11659206.1 \pm 4.1)\times 10^{-10}\, . \\
\end{equation}
Previously there existed a 3.7 $\sigma$ deviation in the SM's estimated value and the measured value by BNL which has now sharpened to 4.2 $\sigma$ considering the FNAL result with a precise deviation of, 

\begin{equation}
\Delta a_{\mu}^{\text{2021}}=(a_{\mu}^{\text{BNL}}+a_{\mu}^{\text{FNAL}})-a_{\mu}^{\text{SM}}= (25.1 \pm 5.9)\times 10^{-10} .\\
\end{equation}

Many new physics scenarios have been proposed so far to explain the anomaly, for an incomplete list of which one may refer~\cite{Jegerlehner:2009ry,Lindner:2016bgg, Majumdar:2020xws,Patra:2016shz} (and references therein) . In this framework new contributions to $a_{\mu}$ arise from the interactions of sterile neutrino $S_L$ due to large mixing with muons. This contribution depends on sterile neutrino mass and its mixing with muons, which are related to light neutrino masses and oscillation parameters in type-II seesaw dominance scheme as discussed earlier. Thus, this scenario opens the possibility of constraning light neutrino masses and mass hierarchy from FNAL results on $a_{\mu}$. Other sizeable contributions can arise from $W_R$ mediated channels and scalar exchange if one considers left-right symmetry breaking at TeV scale. Such cases are studied in one of our recent works \cite{Majumdar:2020xws}, where we have extensively calculated the contributions of neutral fermions, gauge bosons and scalars to $a_{\mu}$ in a TeV scale extended Left-right model.


\section{Conclusion}
\label{sec:concl}
We considered a new mechanism of natural type-II seesaw dominance that allows large light-heavy neutrino mixing within a framework of left-right symmetric model. The fermion sector of the model contains usual quarks and leptons plus an extra sterile neutrino per generation while its scalar sector consists of Higgs doublets, triplets and bidoublet. The extra particles help in generating large light-heavy neutrino mixing that gives new contributions to various 
LFV processes like $\mu \to e \gamma$, $\mu \to 3 e$ and $\mu \to e$ conversion inside nuclei. As a result of these new contributions the new branching ratios can be accessible to present as well as planned experiments. We have demonstrated how the model parameters are suitably adjusted to make the contribution of inverse seesaw and loop induced light neutrino masses sub-dominant. Rather we have generated the light neutrino masses through type-II seesaw mechanism. 

All the physical masses and mixing of neutral leptons are expressed in terms of light neutrino masses and PMNS mixing matrix. Thus the new physics contributions to various LFV processes arising from heavy and sterile neutrinos depend upon light neutrino mass. As a result of this, the bound on absolute scale of light neutrino mass can be derived by saturating the experimental limits on LFV processes, which we find to be in the meV range. We have plotted branching ratios for LFV processes such as $\mu \to e \gamma$, $\mu \to 3 e$ and $\mu \to e$ conversion inside nuclei as a function of the lightest neutrino mass for NH as well as IH pattern of light neutrinos in order to derive the bound on light neutrino mass. We have also plotted dipole moment of electron as a function of light neutrino mass and found that the result depends only upon PMNS phase. Thus in our model the dipole moment of electron can uniquely say about the relative phases of PMNS mixing matrix. We have also studied the new contributions to neutrinoless double beta decay arising from purely left-handed current due to exchange of heavy Majorana neutrinos in our model. It is found that these new contributions can saturate the experimental bound for $m_{1} > 0.001$eV for NH. We have shown the correlation between LFV and LNV  by focusing on $\text{Br}_{\mu\rightarrow e\gamma} \mbox{vs.} T_{1/2}^{0\nu} [^{76}\mbox{Ge}]$ in Fig. \ref{fig:lnvlfv}.
\section*{Acknowledgements}
\label{sec:ackn}
The authors would like to thank P.S. Bhupal Dev for the useful discussions during the early stage of this work.

\newpage
\section{Appendix}
\appendix
\vspace*{-0cm}
\section{Derivation of neutrino masses and mixings in extended left-right seesaw model (ELRSM)}
\label{app:lrsm}
\subsection{ELRSM mass matrix and form of unitary mixing matrix}
We discuss here the implementation of extended seesaw mechanism for neutrino masses and mixing and derivation of type-II seesaw dominance within left-right symmetric models. Apart from usual quarks and leptons, the fermion sector is extended with an extra sterile neutrino $\nu_{S_L}$ (or $\nu_S$ in short-hand notation) while scalar sector consists of usual scalar bidoublet $\Phi$, doublets ($H_R, H_L$) and triplets ($\Delta_R, \Delta_L$). The left-right symmetry is broken down to SM theory through spontaneous symmetry breaking when the scalars $H_R$ or $\Delta_R, \Delta_L$ take VEV. The scalar $H_L$ doesn't play any role here, rather it is present due to the left-right symmetry. The SM theory further breaks down to low energy theory with the help of bidoblet $\Phi$. We call this scheme the extended left-right seesaw model (ELRSM). The neutral fermions neeeded for ELRSM are active left-handed neutrinos, $\nu_L$, active right-handed neutrinos, $\nu_R$ and sterile neutrinos, $\nu_S$. The relevant mass terms within ELRSM are given by 
\begin{eqnarray}
\mathcal{L}_{\rm ELRSM} &=& \mathcal{L}_{M_D}+\mathcal{L}_{M} + \mathcal{L}_{M_{L}} + \mathcal{L}_{M_R} + \mathcal{L}_{\mu_S} \nonumber \\
\mathcal{L}_{M_D} &=& - \sum_{\alpha, \beta} \overline{\nu_{\alpha L}} [M_D]_{\alpha \beta} \nu_{\beta R} \mbox{+ h.c.}\nonumber \\
\mathcal{L}_{M_{}} &=& - \sum_{\alpha, \beta} \overline{\nu_{\alpha S}} [M_{}]_{\alpha \beta} \nu_{\beta R} \mbox{+ h.c.}
\nonumber \\
\mathcal{L}_{M_{L}} &=& - \frac{1}{2} \sum_{\alpha, \beta} \overline{\nu^c_{\alpha L}} [M_{L}]_{\alpha \beta} \nu_{\beta L} \mbox{+ h.c.}
\nonumber \\
\mathcal{L}_{M_{R}} &=& - \frac{1}{2} \sum_{\alpha, \beta} \overline{\nu^c_{\alpha R}} [M_{R}]_{\alpha \beta} \nu_{\beta R} \mbox{+ h.c.}
\nonumber \\
\mathcal{L}_{\mu_{S}} &=& - \frac{1}{2} \sum_{\alpha, \beta} \overline{\nu^c_{\alpha S}} [\mu_{S}]_{\alpha \beta} \nu_{\beta S} \mbox{+ h.c.}
\end{eqnarray}
The flavour states for active left-handed neutrinos $\nu_{\alpha L}$, sterile neutrinos $\nu_{\beta S}$ and right-handed neutrinos $\nu_{\gamma R}$ are defined as follows
\begin{eqnarray}
\nu_{\alpha L} = \begin{pmatrix}
\nu_{eL} \\ \nu_{\mu L} \\ \nu_{\tau L}
\end{pmatrix}\, , \quad 
\nu_{\beta S} = \begin{pmatrix}
\nu_{S_{1 L}} \\ \nu_{S_{2 L}} \\ \nu_{S_{3 L}}
\end{pmatrix}\, , \quad 
\nu_{\gamma R} = \begin{pmatrix}
\nu_{N_{1 R}} \\ \nu_{N_{2 R}} \\ \nu_{N_{3 R}}
\end{pmatrix}\, .
\end{eqnarray}
Similarly, the mass states for neutral fermions are given by
\begin{eqnarray}
\nu_{i L} = \begin{pmatrix}
\nu_{1L} \\ \nu_{2 L} \\ \nu_{3L}
\end{pmatrix}\, , \quad 
S_{j L} = \begin{pmatrix}
S_{1 L} \\ S_{2 L} \\ S_{3 L}
\end{pmatrix}\, , \quad 
N^c_{k R} = \begin{pmatrix}
N^c_{1 R} \\ N^c_{2 R} \\ N^c_{3 R}
\end{pmatrix}\, .
\end{eqnarray}
In the flavour basis $\left( \nu_L, \nu_S, \nu^c_R\right)$ the resulting $9\times 9$ neutral lepton mass matrix within ELRSM is given by,
\begin{eqnarray}
\mathcal{M}_{\rm ELRSM}= \begin{pmatrix}
M_L      & 0           & M_D  \\
0      & \mu_S           & M_{} \\
M^T_D  & M^T_{}    & M_R
\end{pmatrix}
\label{app:elrsm}
\end{eqnarray}
Here $M_D$ is the Dirac mass term connecting $\nu_L-\nu_R$, $M_{}$ is the mixing term between $\nu_R-\nu_S$ while $M_L$, $M_R$, $\mu_S$ are Majorana mass terms for $\nu_L$, $\nu_R$ and $\nu_S$, respectively. The mass hierarchy is given by
\begin{eqnarray}
M_R \gg M_{} > M_D > \mu_S \gg M_L\,.
\label{app:mass-hierarchy}
\end{eqnarray}

The diagonalisation of $\mathcal{M_{\rm ELRSM}}$ after changing it from flavour basis to mass basis is done by a generalized unitary transformation as, 
\begin{eqnarray}
&&\mid \Psi \rangle_{\rm flavor} = \mathcal{V}\mid \Psi \rangle_{\rm mass}\\
&\mbox{or,}& \begin{pmatrix}
\nu_{\alpha L}\\ \nu_{\beta S}\\ \nu^c_{\gamma R}
\end{pmatrix}
= 
\begin{pmatrix}
\mathcal{V}_{\alpha i}^{\nu \nu} & \mathcal{V}_{\alpha j}^{\nu S} & \mathcal{V}_{\alpha k}^{\nu N}\\\mathcal{V}_{\beta i}^{ S \nu} & \mathcal{V}_{\beta j}^{S S} & \mathcal{V}_{\beta k}^{S N}\\\mathcal{V}_{\gamma i}^{N \nu} & \mathcal{V}_{\gamma j}^{N S} & \mathcal{V}_{\gamma k}^{N N}
\end{pmatrix}               
\begin{pmatrix}
\nu_{i}\\ S_{j}\\ N^c_{k}
\end{pmatrix} \, \\
&&   \mathcal{V}^{\dagger} \mathcal{M_{\rm ELRSM}} \mathcal{V}^* 
= \mathcal{\widehat{M}}_{\rm ELRSM} \nonumber \\
&&\hspace*{2.2cm} = \mbox{diag} \left(m_{i},m_{S_j},m_{N_k} \right) \nonumber \\
&&\hspace*{2.2cm} = \mbox{diag} \left(m_{1},m_{2},m_{3},m_{S_1},m_{S_2},m_{S_3},m_{N_1},m_{N_2},m_{N_3} \right)
\end{eqnarray}
Here the indices $\alpha, \beta, \gamma$ run over three generations of light left-handed neutrinos, sterile neutrinos and heavy right-handed neutrinos in flavor basis respectively, whereas the indices $i,j,k$ run over corresponding mass states. 

\subsection{Seesaw block diagonalization for ELRSM neutrino mass matrix} 
\label{app:BD-elrsm}
The block diagonaliztion is done in two steps. In the first step, the ELRSM neutrino mass matrix $\mathcal{M_{\rm ELRSM}}$ is reduced to an `Intermediate Block Diagonal' form $\mathcal{M_{\rm IBD}}$ and in the second step this $\mathcal{M_{\rm IBD}}$ is further diagonalised to $\mathcal{M_{\rm BD}} = \mbox{diag}\left(m_\nu, m_S, m_N \right)$, from which we obtain the mass formulae for three types of neutrinos $\nu_L, \nu_S, \nu_R$ as the three diagonal elements. 
Finally, in order to get the physical masses for all types of neutrinos we need unitary transformations for individual mass matrices as, 
$$\mathcal{M_{\rm diag}} = \mathcal{\widehat{M}}_{\rm ELRSM} =\mbox{diag} \left(m_{i},m_{S_j},m_{N_k} \right)  = \mbox{diag} \left(m_{1},m_{2},m_{3},m_{S_1},m_{S_2},m_{S_3},m_{N_1},m_{N_2},m_{N_3} \right)$$

\subsubsection{Determination of $\mathcal{M_{\rm IBD}}$}
Let us first rewrite the ELRSM mass matrix $\mathcal{M}_{\rm ELRSM}$ given in eq.(\ref{app:elrsm}) to generic type-I+II seesaw~\cite{Grimus:2000vj} as,
\begin{eqnarray}
\mathcal{M_{\rm ELRSM}} &=& \begin{pmatrix}
\mathcal{M}_{L} & \mathcal{M}^T_{D}  \\
\mathcal{M}_{D} & \mathcal{M}_{R}
\end{pmatrix} \, , \nonumber \\
\mbox{where,} && \mathcal{M}_{L}=  \begin{pmatrix}
M_L & 0 \\
0   & \mu_S
\end{pmatrix}_{6 \times 6}
\,, \quad 
\mathcal{M}_{D}=  \begin{pmatrix}
M^T_D & M^T_{}
\end{pmatrix}_{3 \times 6}\, , \quad  
\mathcal{M}_{R} = M_R\,. 
\end{eqnarray}
Using the seesaw approximations given in eq.(\ref{app:mass-hierarchy}), it can be shown that $|\mathcal{M}_{R}| \gg \mathcal{M}_{D} \gg \mathcal{M}_{L}$. First $\mathcal{M_{\rm ELRSM}}$ can be simplified to intermediate block diagonalized form $\mathcal{M_{\rm IBD}} $ by integrating out the heaviest right-handed neutrinos from other neutral states. Thus, the first block diagonalized approximate unitary mixing matrix $\mathcal{W}_1$ gives 
\begin{eqnarray}
&&\mathcal{W}^T_1 \mathcal{M}^{}_{\rm ELRSM} \mathcal{W}_1=\mathcal{M}^{}_{\rm IBD} \nonumber \\
\text{where},
&& \mathcal{M}^{}_{\rm IBD} = \begin{pmatrix}
M^{\rm light}_{\rm Eff}        & {\bf 0}_{3 \times 6} \\
{\bf 0}_{3 \times 6}  & M^{\rm heavy}_{}
\end{pmatrix}, \nonumber \\
\text{and}
&&\mathcal{W}_1=\begin{pmatrix}
\sqrt{1-\mathcal{B}\mathcal{B}^{\dagger}} & \mathcal{B}\\
-\mathcal{B}^{\dagger} & \sqrt{1-\mathcal{B}^{\dagger}\mathcal{B}}
\end{pmatrix}
\end{eqnarray}
Now $\mathcal{M_{\rm IBD}} $ gives effective mass matrix in the modified basis of left-handed active and sterile neutrinos as,
\begin{eqnarray}
M^{\rm light}_{\rm Eff} &\equiv& M^{\rm Eff}_{ELRSM} =  \mathcal{M}_{L}- \mathcal{M}^T_{D} \mathcal{M}^{-1}_{R} \mathcal{M}_{D} \nonumber \\
&=& 
\begin{pmatrix}
M_L & 0 \\
0   & \mu_S
\end{pmatrix} -  \begin{pmatrix}
M_D \\
M_{}
\end{pmatrix} M^{-1}_R
\begin{pmatrix}
M^T_D & M^T_{} 
\end{pmatrix} \nonumber \\
&=& \begin{pmatrix}
M_L & 0 \\
0   & \mu_S
\end{pmatrix} - \begin{pmatrix}
M_{D} M^{-1}_R M^T_{D}    & M_{D} M^{-1}_R M^T_{}  \\
M_{} M^{-1}_R M_{D}          &  M_{} M^{-1}_R M^T_{}
\end{pmatrix} \nonumber \\
&=& \begin{pmatrix}
M_L-M_{D} M^{-1}_R M^T_{D}    & - M_{D} M^{-1}_R M^T_{}  \\
-M_{} M^{-1}_R M_{D}          & \mu_S - M_{} M^{-1}_R M^T_{}
\end{pmatrix}
\label{eq:nuL-nuS}
\end{eqnarray}
and block diagonalized mass formula for the integrated out heavy right-handed neutrinos as,
\begin{eqnarray}
M^{\rm heavy}_{}&\equiv& m_N =
M_R + \cdots
\label{eq:nuR}
\end{eqnarray}
Using the standard seesaw block diagonalization methodology, one can get 
$$\mathcal{B}^\dagger_1 = \begin{pmatrix}
M_D M^{-1}_R \\ M M^{-1}_R
\end{pmatrix}^T
$$and the approximated intermediate block diagonalized mixing matrix up to the order of $\mathcal{O}(1/M_R)$ is given by
\begin{eqnarray}
\mathcal{W}_1 &=& \begin{pmatrix}
1- \frac{1}{2} Z Z^\dagger & -\frac{1}{2} Z Y^\dagger & Z  \\
-\frac{1}{2} Y Z^\dagger & 1- \frac{1}{2} Y Y^\dagger  & Y  \\
- Z^\dagger   & -Y^\dagger &  1-\frac{1}{2}\left(Z^\dagger Z + Y^\dagger Y\right)
\end{pmatrix}
\end{eqnarray}
where $Z=M_D M^{-1}_R$ and $Y=M_{} M^{-1}_{R}$. 

In this way the light active and sterile neutrinos contained in the effective block diagonalized mass matrix $M^{\rm light}_{\rm Eff}$ get completely decoupled from heaviest right-handed neutrinos. Thus the first seesaw intermediate block diagonalization brings down a $9 \times 9$ matrix into two smaller matrices; a block diagonalized $6\times 6$ matrix for $\nu_L$ and $\nu_S$ and a $3\times 3$ matrix for $\nu_R$. Now we have to repeat the same seesaw block diagonalization procedure for $M^{\rm light}_{\rm Eff}$ to further block diagonalize the light neutrino states. 

\subsubsection{Determination of $\mathcal{M_{\rm BD}}$}
We require another approximated unitary mixing matrix $\mathcal{S}$ (and $\mathcal{W}_2$ for accounting the integrated out right-handed neutrinos) to further block diagonalize the effective light neutrino mass matrix $M^{\rm light}_{\rm Eff}$ in order to get mass matrices for $\nu_L$ and $\nu_S$. Let us write $M^{\rm light}_{\rm Eff}$ as given in eq.\ref{eq:nuL-nuS} in a simpler form as,
\begin{eqnarray}
M^{\rm light}_{\rm Eff} &=&-
\begin{pmatrix}
M_L - M_{D} M^{-1}_R M^T_{D}    & - M_{D} M^{-1}_R M^T_{}  \\
-M_{} M^{-1}_R M_{D}          &  \mu_S - M_{} M^{-1}_R M^T_{}
\end{pmatrix} = 
\begin{pmatrix}
\mathcal{M}^\prime_L& \mathcal{M}^{\prime T}_D  \\
\mathcal{M}^\prime_{D} & \mathcal{M}^\prime_{R}
\end{pmatrix} \, , \nonumber \\
\mbox{where,} && \mathcal{M}^\prime_{L}= M_L-M_{D} M^{-1}_R M^T_{D} \nonumber \\
&&            \mathcal{M}^\prime_{D}= -M_{} M^{-1}_R M^T_{D} \nonumber \\
&& \mathcal{M}^\prime_{R}=\mu_S -M_{} M^{-1}_R M^T_{}
\label{eq:nuL-nuS-b}
\end{eqnarray}
Now repeating the same procedure of type-I+II seesaw block diagonalization along with the mass hierarchy given in eq.(\ref{app:mass-hierarchy}), the diagonalized mass matrix and mixing matrix look as follows,
\begin{eqnarray}
&&\mathcal{S}^T \mathcal{M}^{\rm light}_{Eff} \mathcal{S}
=\begin{pmatrix}
m_\nu & 0 \\
0     & m_S
\end{pmatrix}
\nonumber \\
&& \mathcal{S}=\begin{pmatrix}
\sqrt{1-\mathcal{A}\mathcal{A}^{\dagger}} & \mathcal{A}\\
-\mathcal{A}^{\dagger} & \sqrt{1-\mathcal{A}^{\dagger}\mathcal{A}}
\end{pmatrix}
=\begin{pmatrix}
1- \frac{1}{2} X X^\dagger & X \\
-X^\dagger & 1- \frac{1}{2} X^\dagger X
\end{pmatrix}
\end{eqnarray}
where active-sterile neutrino mixing is given by the matrix $X=M_D M^{-1}$. Thus, we get the block diagonalized mass formulae for light active neutrinos, sterile neutrinos and right-handed neutrinos as follows.
\begin{eqnarray}
&&\mathcal{W}^T_2 \mathcal{M}^{IBD}_{\rm ELRSM} \mathcal{W}_2=\mathcal{M}^{BD}_{\rm ELRSM} \nonumber \\
&& \mathcal{M}^{\rm BD}_{\rm ELRSM} = \begin{pmatrix}
m_\nu  &  0        & 0\\
0      &  m_S      & 0 \\
0      &  0        & m_N
\end{pmatrix} 
\end{eqnarray}

The approximated unitary mixing matrix $\mathcal{W}_2$ is given by,
\begin{eqnarray}
&& \mathcal{W}_2 = 
\begin{pmatrix}
\mathcal{S}  & {\bf 0}_{6 \times 3} \\
{\bf 0}_{3 \times 6} & {\bf 1}_{3 \times 3}
\end{pmatrix} \nonumber \\
&&\hspace*{1cm} =\begin{pmatrix}
\sqrt{1-\mathcal{A}\mathcal{A}^{\dagger}} & \mathcal{A}   &  {\bf 0}_{3 \times 3} \\
-\mathcal{A}^{\dagger} & \sqrt{1-\mathcal{A}^{\dagger}\mathcal{A}} & {\bf 0}_{3 \times 3}\\
{\bf 0}_{3 \times 3}  & {\bf 0}_{3 \times 3} & {\bf 1}_{3 \times 3} 
\end{pmatrix}
=\begin{pmatrix}
1- \frac{1}{2} X X^\dagger & X  &  {\bf 0}_{3 \times 3} \\
-X^\dagger & 1- \frac{1}{2} X^\dagger X & {\bf 0}_{3 \times 3}\\
{\bf 0}_{3 \times 3}  & {\bf 0}_{3 \times 3} & {\bf 1}_{3 \times 3} 
\end{pmatrix}
\end{eqnarray}

\subsubsection{Radiative contribution to light neutrino masses}
The block diagonalized mass formulas for light active neutrinos, sterile neutrinos and heavy neutrinos are given by
\begin{eqnarray}
m_{\nu} &=& \mathcal{M}^\prime_{L}- \mathcal{M}^{\prime T}_{D} \mathcal{M}^{\prime^{-1}}_{R} \mathcal{M}^{\prime}_{D}\nonumber \\
&=&M_L - M_{D} M^{-1}_R M^T_{D} - 
\big(-M_{D} M^{-1}_R M^T_{} \big) 
\cdot \big(-M_{} M^{-1}_R M^T_{} \big)^{-1} 
\cdot \big(\mu_S -M_{} M^{-1}_R M^T_{D} \big) \nonumber \\
&=&M_L-M_{D} M^{-1}_R M^T_{D} + M_{D} M^{-1}_R M^T_{D}
+ \big(\frac{M_D}{M}\big) \mu_S \big(\frac{M_D}{M}\big)^T  \nonumber \\
&=& m^{\rm II} + m^{\rm inv}       \label{eq:mnu} \\
m_{S} &=& \mu_S -M_{} M^{-1}_R M^T_{}  \\
\label{eq:mS}
m_{N} &=& M_R 
\label{eq:mN}
\end{eqnarray}
An interesting feature of approximated seesaw block diagonalization scheme within ELRSM is that   type-I seesaw contribution gets exactly cancelled out at tree level. Thus at tree level light neutrinos get mass through type-II seesaw and inverse seesaw. There is also a sizable contribution to light neutrino masses at 1-loop level.
\begin{eqnarray}
m_\nu = m^{\rm II} + m^{\rm inv} + m^{rad}_\nu \,.
\end{eqnarray}
The analytic expression for this one loop contribution to light neutrino mass mediated by SM $W$ and $Z$-bosons is given by
\begin{eqnarray}
m^{\rm rad}_\nu \equiv \Delta M &\simeq & M_D \frac{\alpha_W}{16\pi
	m_W^2} M_R\left[\frac{m_H^2}{M_R^2-m_H^2{\bf 1}_3}\ln
\left(\frac{M_R^2}{m_H^2}\right) +   
\frac{3m_Z^2}{ M_R^2-m_Z^2{\bf 1}_3}\ln\left(\frac{
	M_R^2}{m_Z^2}\right)\right] M^T_D\nonumber\\ 
& \simeq &  
M_D M_R^{-1} x_R\, f(x_R) M_D^{T}
\end{eqnarray}
where the one-loop function $f(x_R)$ is defined as 
\begin{eqnarray}
f(x_R) =
\frac{\alpha_W}{16\pi}\left[\frac{x_H}{x_R-x_H}\ln\left(\frac{x_R}{x_H}\right)
+ \frac{3x_Z}{x_R-x_Z}\ln\left(\frac{x_R}{x_Z}\right) \right] 
\end{eqnarray} 
with   $x_R   \equiv\hat{M}_R^2/m_W^2$,   $x_H\equiv   m_H^2/m_W^2$,
$x_Z\equiv m_Z^2/m_W^2$, $\hat{M}_R$ as diagonal matrix.

\subsubsection{Complete diagonalization and physical neutrino masses}
After block diagonalization, the mass matrix for the three types of neutrinos are further diagonalized by respective unitary mixing matrices as follows,
\begin{eqnarray}
\mathcal{U} = \begin{pmatrix}
U_\nu & 0 & 0 \\
0   & U_S  & 0 \\
0  &  0 &  U_N
\end{pmatrix}
\end{eqnarray}
resulting in physical masses for all the neutrinos.
\begin{eqnarray}
&&U^\dagger_\nu m_\nu U^*_\nu = \hat{m}_\nu = \mbox{diag}\left(m_{\nu_1}, m_{\nu_2}, m_{\nu_3} \right) \nonumber \\
&&U^\dagger_S m_S U^*_S = \hat{m}_S = \mbox{diag}\left(m_{s_1}, m_{s_2}, m_{s_3} \right) \nonumber \\
&&U^\dagger_N m_N U^*_N = \hat{m}_N = \mbox{diag}\left(m_{N_1}, m_{N_2}, m_{N_3} \right)
\end{eqnarray}
\begin{eqnarray}
\mathcal{U}_{9 \times 9} = \begin{pmatrix}
{\bf U_\nu}_{3 \times 3}  &  {\bf 0}_{3 \times 3}  & {\bf 0}_{3 \times 3}  \\
{\bf 0}_{3 \times 3}  &  {U_S}_{3 \times 3}    & {\bf 0}_{3 \times 3}  \\
{\bf 0}_{3 \times 3}  &  {\bf 0}_{3 \times 3}  & {U_N}_{3 \times 3}
\end{pmatrix} 
\end{eqnarray}

The complete $9\times 9$ mixing matrix is derived as,
\begin{eqnarray}
\mathcal{V} &=& \mathcal{W}_1 \cdot \mathcal{W}_2 \cdot \mathcal{U} \nonumber \\
&&\hspace*{-0.5cm}=\begin{pmatrix}
1 -\frac{1}{2} Z Z^\dagger   &  -\frac{1}{2} Z Y^\dagger  &  Z \\
-\frac{1}{2} Y Z^\dagger  &  1-\frac{1}{2} Y Y^\dagger  & Y  \\
-Z^\dagger   & -Y^\dagger  & 1-\frac{1}{2}\left(Z^\dagger Z + Y^\dagger Y \right)
\end{pmatrix} \cdot 
\begin{pmatrix}
1 -\frac{1}{2} X X^\dagger & X   & 0 \\
-X^\dagger & 1 -\frac{1}{2} X^\dagger X  & 0 \\
0    &     0       & 1   
\end{pmatrix} \cdot 
\begin{pmatrix}
U_\nu & 0 & 0 \\
0   & U_S  & 0 \\
0  &  0 &  U_N
\end{pmatrix} \nonumber \\
&\simeq& \begin{pmatrix}
U_{\nu}\left(1-\frac{1}{2}X X^{\dagger} \right)  & -U_{S}X &  Z U_N\\
-U_{\nu}X^{\dagger}  & U_S \left(1-\frac{1}{2}X^{\dagger}X \right) & U_{N} Y \\
\left(Z^\dagger X^\dagger X\right)U_\nu & -U_S Y^{\dagger} & U_N\left(1-\frac{1}{2}Y^{\dagger}Y \right)
\end{pmatrix}
\end{eqnarray}
Putting $ X = M_{D} M^{-1}_{}$, $Y = M_{} M^{-1}_{R}$, $Z=M_D M^{-1}_R$ and fixing the typical magnitudes for $ M_D \simeq $~10\,GeV ,$ M_{} \simeq $~100\,GeV, $M_{R} \simeq$ 1000\,GeV - 10\,TeV, we get $X \simeq 0.1 $, $Y\simeq 0.01$, $Z\simeq 10^{-3}$.  Since $U_{\nu}$, $U_N$ and $U_S$ are of $\mathcal{O}(1)$, the matrix elements of $\mathcal{V}$ are approximated to be
\begin{eqnarray}
\begin{pmatrix}
\mathcal{V}_{\alpha i}^{\nu \nu} & \mathcal{V}_{\alpha j}^{\nu S} & \mathcal{V}_{\alpha k}^{\nu N}\\\mathcal{V}_{\beta i}^{ S \nu} & \mathcal{V}_{\beta j}^{S S} & \mathcal{V}_{\beta k}^{S N}\\\mathcal{V}_{\gamma i}^{N \nu} & \mathcal{V}_{\gamma j}^{N S} & \mathcal{V}_{\gamma k}^{N N}
\end{pmatrix}  \simeq \begin{pmatrix}
1       & 0.1     & 10^{-3} \\
0.1     & 0.91    & 0.01   \\
0       & 0.01   & 1.0
\end{pmatrix}
\end{eqnarray} 

\subsection{Achieving type-II seesaw dominance by including radiative contributions}
In addition to the generic Dirac neutrino mass $M_D$ and Majorana neutrino mass $M_R$, there are two more terms; a mixing term $M$ that connects right-handed neutrinos with extra serile neutrinos and a small Majorana mass term $\mu_S$ for only sterile neutrinos. With a small but non-zero $\mu_S$, an additional contribution to light neutrino masses arises through inverse seesaw mechanism as $m^{\rm inv}_{\nu} = (M_D/M) \mu_S (M_D/M)^T$ apart from the type-II seesaw term at tree level. This inverse seesaw term can be avoided and type-II seesaw dominance can be achieved with $\mu_S \to 0$. However SM radiative corrections involving the SM $Z$ and Higgs boson also contribute to light neutrino masses $m^{\rm rad}_{\nu}$ as pointed out originally by \cite{Pilaftsis:1991ug} and later by \cite{Dev:2012bd,Dev:2012sg}. The presence of inverse seesaw and one loop radiative corrections to light neutrino mass might spoil the plot for type-II seesaw dominance. We briefly discuss below how one can obtain the type-II seesaw dominance.
\begin{itemize}
 \item {\bf By considering suppressed value of Dirac neutrino mass:-}\\
 The total contribution to light neutrino masses is the sum of type-II, inverse and radiative seesaw mechanism, out of which the inverse and radiative seesaw contributions mostly depend on Dirac neutrino masses. Thus a suppressed value of Dirac neutrino mass will give negligible contribution to inverse as well as radiative corrections to light neutrinos masses. In such case (i.e. in the limit of suppressed or vanishing Dirac neutrino masses), the only scheme for generating light neutrino masses will be type-II  seesaw dominance. 
 \item {\bf Due to effective cancellation between inverse and radiative contributions:-}\\
 Alternatively type-II seesaw dominance can be achieved by allowing exact/partial cancellation between inverse seesaw and radiative contribution such that they marginally/sub-dominantly contribute to the light neutrino masses. This is phenomenologically more interesting because it allows large light-heavy neutrino mixing which can contribute to lepton number violation and lepton flavour violation. We have followed this method in our present model, i.e. we have allowed cancellation to make $m_\nu^{inv}+m_\nu^{\rm rad} = 0$.
\end{itemize}

\section{Loop functions involved in lepton flavour violating processes} 
\label{sec:form_loop}
The relevant loop functions arising in various lepton flavour violating (LFV) processes 
like $\mu \to e \gamma$, $\mu \to 3 e$ and $\mu \to e$ conversion inside a nuclei used in our analysis are 
given by
\begin{align} \label{eq:loop_funcs}
\begin{split}
\mathcal{G}^{\gamma}_1(x) &= -\frac{2x^3+5x^2-x}{4(1-x)^3} - \frac{3x^3}{2(1-x)^4}\ln{x}, \\[1mm]
\mathcal{G}^{\gamma}_2(x) &= \frac{x^2-11 x+4}{2 (1-x)^2}-\frac{3 x^2}{(1-x)^3}\ln{x}, \\[1mm]
\mathcal{F}_{\gamma}(x) &= \frac{7x^3-x^2-12x}{12(1-x)^3}-\frac{x^4-10x^3+12x^2}{6(1-x)^4}\ln{x}, \\[1mm]
\end{split} 
\end{align}


\begin{align} \label{eq:loop_funcs}
\begin{split}
\mathcal{F}_{\rm box}(x,y) &= \left(4+\frac{xy}{4}\right)I_2(x,y,1)-2xyI_1(x,y,1), \\[1mm]
\mathcal{F}_{\rm X box}(x,y) &= -\left(1+\frac{xy}{4}\right)I_2(x,y,1)-2xyI_1(x,y,1), \\[1mm] 
\mathcal{G}_{\rm box}(x,y,\eta) &= -\sqrt{x y} \left[(4+x y \eta) I_1(x,y,\eta)-(1+\eta)I_2(x,y,\eta)\right], \\
\mathcal{G}_Z(0,x) &= -\frac{x\ln{x}}{2(1-x)}\,, \\[1mm]
\mathcal{F}_{\rm box}(0,x) &= \frac{4}{1-x}+\frac{4x}{(1-x)^2}\ln{x}\, , \\[1mm]
\mathcal{F}_{\rm X box}(0,x) &= -\frac{1}{1-x}-\frac{x\ln{x}}{(1-x)^2}\,, \\[1mm]
\mathcal{F}_Z(x) &= -\frac{5x}{2(1-x)}-\frac{5x^2}{2(1-x)^2}\ln{x}\,, \\[1mm] 
\end{split} 
\end{align}
These loop functions involve heavy right-handed neutrino mass $m_N$ and sterile neutrino mass $m_S$. In our model these masses $m_N$ and $m_S$ are proportional to light neutrino mass $m_\nu$. Thus the relevant loop functions needed for various LFV processes can be expressed in terms of lightest neutrino mass.

\section{Expression for $\mu\longrightarrow e$ conversion with the mediation of neutrinos $N$ and $S$}
\label{sec:mutoe}
\begin{equation}
{\rm R}^{A(N,Z)}_{\mue} = \frac{\alpha_{\rm em}^3\alpha_W^4m_\mu^5}{16\pi^2\mwl^4\Gamma_{\rm capt}}
\frac{Z_{\rm eff}^4}{Z}\left|\mathcal{F}(-m_\mu^2)\right|^2 
\left(|\mathcal{Q}_L^W|^2+|\mathcal{Q}_R^W|^2\right), \label{eq:brmue_full}
\end{equation}
where, the relevant terms containing only light-heavy neutrino mixing are given by
\begin{eqnarray}
&& \mathcal{Q}^W_{L} = (2 Z+N)\left[\mathcal{W}^u_{L}-\frac{2}{3} s_W^2 G^\gamma_{R}\right] 
+ (Z+2N)\left[W^d_{L} +\frac{1}{3}s_W^2 G^\gamma_{R} \right],\nonumber \\
&& \mathcal{Q}^W_{R} = (2 Z+N)\left[\mathcal{W}^u_{R}\right] + (Z+2N)\left[W^d_{R}\right],
\end{eqnarray}
and the factor used here are expressed as,
\begin{eqnarray}
&&W^u_{L}= \frac{2}{3} s_W^2 \mathcal{F}^\gamma_{L} + \left(-\frac{1}{4}+\frac{2}{3} s_W^2\right)\mathcal{F}^{Z_1}_{L}
+\frac{1}{4}\mathcal{B}^{\mu e uu}_{LL}, \nonumber \\ 
&&W^u_{R}\to 0 , \nonumber \\ 
&&W^d_{L} = -\frac{1}{3} s^2_W \mathcal{F}^\gamma_{L} + \left(\frac{1}{4}-\frac{1}{3} s_W^2\right) \mathcal{F}^{Z_1}_{L} 
+\frac{1}{4}\mathcal{B}^{\mu e dd}_{LL}, \nonumber \\
&&W^d_{R} \to 0 \,.
\end{eqnarray}

\begin{equation}
\frac{V^{(p)}}{\sqrt{Z}} = \frac{Z_{eff}^2F(-m_\mu^2)\alpha_{\rm em}^{\frac{3}{2}}}{4\pi}\, ,
\end{equation}
and $V^{(p)}/Z \simeq V^{(n)}/N$. The key box diagram form factors are expressed as
\begin{align}
\begin{split}
\mathcal{B}^{\mu e uu}_{LL} &= \sum_{i=1}^3 \bigg\{{\mbox{V}^{\nu N}_{\mu i}}^* {\mbox{V}^{\nu N}_{e i}} 
\left[\mathcal{F}_{\rm box}(0,x_i) - \mathcal{F}_{\rm box}(0,0)\right], \\
\mathcal{B}^{\mu e dd}_{LL} &\simeq \sum_{i=1}^3 {\mbox{V}^{\nu N}_{\mu i}}^* {\mbox{V}^{\nu N}_{e i}}  
\left\{\mathcal{F}_{\rm X box}(0,x_i) - \mathcal{F}_{\rm X box}(0,0)  \right. \\ 
& \left. + |V_{td}|^2\left[\mathcal{F}_{\rm X box}(x_t,x_i)-\mathcal{F}_{\rm X box}(0,x_i)-\mathcal{F}_{\rm X box}(0,x_t)
+F_{\rm X box}(0,0)\right]\right\}, \\
\mathcal{B}^{\mu e qq}_{RR} &= \frac{\mwl^2}{\mwr^2} \mathcal{B}^{\mu e qq}_{LL}(\mbox{V}^{\nu N} 
\leftrightarrow {\mbox{V}^{N N}}^*\,;\,x^N_i \leftrightarrow y^N_i\,;\,x_t \leftrightarrow y_t) \to 0\, ,
\end{split}
\end{align}
where $x_t = m_t^2/\mwl^2$ and $y_t = m_t^2/\mwr^2$.

 \bibliographystyle{utcaps_mod}
\bibliography{onubb_LR}
\end{document}